\documentclass{ieeeaccess}
\usepackage{cite}
\usepackage{amsmath,amssymb,amsfonts}
\usepackage{algorithmic}
\usepackage{graphicx}
\usepackage{textcomp}
\usepackage{amssymb}
\usepackage{latexsym}
\usepackage{url}
\usepackage{times,amsmath}
\usepackage{stfloats}
\usepackage{graphicx}
\usepackage{subfig}
\usepackage[titles,subfigure]{tocloft}
\usepackage{makeidx} 
\usepackage{lineno}
\usepackage{array}
\usepackage[percent]{overpic}
\usepackage{amsfonts}
\usepackage{eurosym}
\usepackage{multirow}
\usepackage{mathtools}
\usepackage{booktabs} 
\usepackage{pdfpages}
\def\BibTeX{{\rm B\kern-.05em{\sc i\kern-.025em b}\kern-.08em
    T\kern-.1667em\lower.7ex\hbox{E}\kern-.125emX}}
\begin{document}
\history{Date of publication xxxx 00, 0000, date of current version xxxx 00, 0000.}
\doi{10.1109/ACCESS.2017.DOI}

\title{Unified Model Selection Approach Based on Minimum Description Length Principle in Granger Causality Analysis}
\author{\uppercase{Fei Li}\authorrefmark{1},  \uppercase{Xuewei Wang}\authorrefmark{1}, \uppercase{Qiang Lin\authorrefmark{1}, and Zhenghui Hu}\authorrefmark{1}}
\address[1]{College of Science, Zhejiang University of Technology, Hangzhou, Zhejiang 310023, China.}

\tfootnote{This work is supported in part by National Key R$ \& $D Program of China under Grant 2018YFA0701400, in part by the Public Projects of Science Technology Department of Zhejiang Province under Grant LGF20H180015.}

\markboth
{Author \headeretal: Preparation of Papers for IEEE TRANSACTIONS and JOURNALS}
{Author \headeretal: Preparation of Papers for IEEE TRANSACTIONS and JOURNALS}

\corresp{Corresponding author: Zhenghui Hu (e-mail: zhenghui@zjut.edu.cn).}

\begin{abstract}
Granger causality analysis (GCA) provides a powerful tool for uncovering the patterns of brain connectivity mechanism using neuroimaging techniques. 
In this paper, distinct from conventional two-stage GCA, we present a unified model selection approach based on the minimum description length (MDL) principle for GCA in the context of the general regression model paradigm. In comparison with conventional methods, our approach emphasize that model selection should follow a single mathematical theory during the GCA process. Under this framework, all candidate models within the model space might be compared freely in the context of the code length, without the need for an intermediate model. We illustrate its advantages over conventional two-stage GCA approach in a 3-node network and a 5-node network synthetic experiments. The unified model selection approach is capable of identifying the actual connectivity while avoiding the false influences of noise. More importantly, the proposed approach obtained more consistent results in a challenging fMRI dataset, in which visual/auditory stimulus with the same presentation design gives identical neural correlates of mental calculation, allowing one to evaluate the performance of different GCA methods. Moreover, the proposed approach has potential to accommodate other Granger causality representations in other function space. The comparison between different GC representations in different function spaces can also be naturally deal with in the framework.
\end{abstract}


\titlepgskip=-15pt

\maketitle

\section{Introduction}
\label{sec:intro}

\PARstart{C}{ausal} connectivity analysis, also called effective connectivity, plays an increasingly important role in brain research using neuroimaging techniques \cite{Amunts2016The,Martin2016The,Poo2016China,Committee2016Australian,Okano2016Brain,Jeong2016Korea}. It reflects a trend in neuroscience away from focusing on individual brain unit (functional specialization) toward complex neural circuits (functional integration) at different spatial scales \cite{strogatz2001exploring,Sporns2004Organization,Bullmore2009Complex}, where the integration among specialized areas is mediated by causal connectivity \cite{bressler2010large,Rubinov2010Complex,sporns2011human,Siegel2012Spectral,Hutchison_NeuroImage13,Valk2017Socio}. Causal connectivity refers to the influence that one neural system exerts over another, either in the absence of identifiable behavioral events or in the context of task performance \cite{Stephan_NeuroImage12,Tsunada_NN16}. It provides important insights into brain organization. Causal connectivity analysis is based on temporal relations existing between time series recordings of neural activity, which may be obtained by neuroimaging techniques, such as electroencephalography (EEG) \cite{Olivier2008Identifying,van2014alpha,Bastos2015Visual}, local field potentials (LFP) \cite{brovelli2004beta,Wang2008Estimating,Bastos2015Visual}, magnetoencephalography (MEG) \cite{gow2008lexical,Ploner2010Functional,florin2010effect}, and functional magnetic resonance imaging (fMRI) \cite{Roebroeck2005Mapping,sridharan2008critical,Menon2010saliency,Ryali2011Multivariate} etc.

Granger causality analysis (GCA) is an effective tool for detecting the causal connections that can provide information about the dynamics and directionality of the associations. Causality in the Granger sense is based on the statistical predictability of one time series that derives from knowledge of one or more other time series \cite{wiener1956theory,Granger1969Investigating}. For two time series, $X$ and $Y$, that are stochastic and wide-sense stationary (i.e., constant means and variances) \cite{Granger_JE74}, the general idea of Granger causality is that variable $Y$ is said to have a causal influence on variable $X$ when the prediction of variable $X$ could be improved by incorporating the history information of variable $Y$. 

In practical applications, the GCA is usually conveyed in the context of linear autoregressive (AR) models of stochastic processes. The AR models with different time lags represent the historical dependence on variables, with or without including other variables, to forecast the dynamics of the variable. The optimal model for predication then is chosen by model selection techniques. The extension of GCA to nonlinear model that capture nonlinear causal relations is also available \cite{chavez2003statistical,gourevitch2006linear}. And the validity of original GCA can also be affected if the errors after the prediction with the model is not normally distributed \cite{Goebel_MRI03,Roebroeck2005Mapping}. Several test methods of causality in Granger sense for the data with asymptotical noise distribution have been developed that is especially useful for task-related fMRI studies \cite{Hacker_AE06,Kaminski}. Asymmetric causality testing recently has also been suggested in order to separate the causal impacts of positive and negative changes \cite{Hatemi-J_EE12}. Moreover, Granger causality have been also expressed in other function spaces, e.g. Fourier spaces (frequency domain) \cite{John1982Measurement,brovelli2004beta,Ding2006Granger,Guo2008Uncovering,seth2010matlab,Barnett2014The,Stokes2017A,ning2017dynamic}, kernel Hilbert spaces \cite{liao2009kernel,Chen_NC14}. These methods are important in neuroscience studies since the causal influences between neuronal populations are often nonlinear and have complicated statistics due to various sources of uncertainties. On the other hand, the original GCA is limited to investigate causal connectivity between two nodes. There have been numerous efforts on expanding the approach from small network with few nodes to large, complex network \cite{Newman_PRE04,Marinazzo_PRL08}. A recent approach presented an iterative scheme that gradually pruned the network by removing indirect connections, considered one at a time, to uncover large network structure with hundreds of nodes \cite{Zou_BMCB10}.

Despite these developments, the basic idea of Granger causality remains unchanged. It generally comprises two stages in this framework: ($1$) specifying the model order (the number of time lags that is associated with its own history information and external effects), and ($2$) deciding the optimal model. The order of the predictors is usually determined with the Akaike information criterion (AIC) \cite{Akaike1974new} or the Bayesian information criterion (BIC) \cite{Schwarz1978estimating}, whereas the optimal predictor is judged through statistical pairwise comparison \cite{Deshpande2010Multivariate,Guo2010Granger,Bressler2011Wiener,Li2010A}. Specifying model order and selecting the optimal model comply two completely different mathematical theories.

Indeed, from a mathematical perspective, the two stages are all to select the best model capturing essential features under investigation from a number of competing models in terms of given observations. They are essentially the same problem of model selection. Model selection is the most important aspect of inference with causal models and allows one to test different hypotheses by comparing different models in terms of selection criteria. Two different theories applied in the same question might generate different selection criteria and therefore degrade the performance of GCA. 

We argue that the two stages in GCA are the problem of model selection, and should obey the same benchmark under one mathematical theory. In this paper, against two-stage selection scheme, we therefore proposed a unify model selection approach for GCA with the minimum description length (MDL) principle that model selection should follow a single mathematical theory during the GCA process. We illustrated the benefits of introducing a unified model selection approach in simulated and real fMRI experiments. Especially, we compared the proposed approach with conventional two-stage GCA one in an empirical fMRI dataset for the validation of causal connectivity analysis.

The rest of the article is organized as follows. Section \ref{sec:problem statement} introduces the basic GCA concept and discusses the potential problems in the current two-stage GCA scheme and our motivation for introducing the MDL principle in GCA. In Section \ref{sec:the MDL principle}, the MDL principle has been illustrated in detail, and the formula of two-part MDL also has been derived with Bernoulli distribution in Markov model class. Immediately, in Section \ref{sec:MDL guided in GCA}, the MDL guided model selection for linear model has been carried out, in time domain and frequency domain respectively. In Section \ref{sec:Experimental}, We illustrate the advantages of our proposal over conventional two-stage GCA approach in a 3-node network and a 5-node network synthetic experiments. At the same time, the proposed approach obtained more consistent results in a challenge fMRI dataset for causality investigation, mental calculation network under visual and auditory stimulus, respectively. Section \ref{sec:conclusion} demonstrates the comparison between conventional two-stage GCA and our proposal from modeling standpoint, and discusses its potential development in a wider Granger causality sense.

\section{Problem Statement}
\label{sec:problem statement}

Let variables $X$ and $Y$ be two stochastic and stationary time series. Now consider the following pair of (restricted and unrestricted) regression models 
\begin{equation}
\begin{cases}
X[n] = \sum_{i = 1}^p {{\beta _x}[i]X[n - i]}  + u[n]\\
X[n] = \sum_{i = 1}^p {{\beta _x}[i]X[n - i]}  + \sum_{i = 1}^q {{\beta _y}[i]Y[n - i]}  + v[n],
\end{cases}
\label{eqn:restricted and unrestricted}
\end{equation}
where $p$ and $q$ are the model orders (the numbers of time lags) in $X$ and $Y$, respectively, $\beta_x$ and $\beta_y$ are the model coefficients, and $u$ and $v$ are the residual of the models. The order of historical predictor $p$ is usually determined with the AIC \cite{Akaike1974new}  or the BIC \cite{Schwarz1978estimating},
$$AIC=-2log(L(\theta))+2k $$ 
$$BIC=-2log(L(\theta))+klog(n)$$
Where $ n $ is the sample size, and k is the number of parameters which your model estimates, and $ \theta $ is the set of all parameters. $ L(\theta) $ represents the likelihood of the model tested, given your data, when evaluated at maximum likelihood values of $ \theta $. 

The conventional GCA requires statistical significance to determine whether the unrestricted model provides a better prediction than the restricted model. The hierarchical $F$-statistics, based on the \textsl{extra sum-of-squares} principle \cite{Casella}, can be used to evaluate significant predictability, given as 
\begin{equation}
F=\dfrac{(RSS_{r}-RSS_{u})/q}{RSS_{u}/(n-p-q-1)}\quad vs \quad F_{0}(q,n-p-q-1)
\label{eqn:F-statistics}
\end{equation}
where $ RSS_{r} $ and $ RSS_{u} $ represent the sum of squared residuals of restricted model and unrestricted model, respectively, $n$ is the total number of observations to estimate the unrestricted model. The $F$-statistics approximately follows an $F$-distribution with ($q$, $n-p-q-1$) degrees of freedom. 

If  $ F \textgreater F_{0}(q,n-p-q-1) $, the variability of the residual of unrestricted model is significantly less than the variability of the residual of restricted model, then there is an improvement in the prediction of $X$ due to $Y$, and we refer to this as causal influence from $Y$ to $X$. But in the current bivariate model, spurious connections will emerge frequently. 

In order to remove spurious connections caused by indirect causalities between nodes, GCA also provides a measure of conditional causal connection by introducing another variable $Z$ into Eqs. (\ref{eqn:restricted and unrestricted}):
\begin{equation}
\begin{cases}
X[n] = \sum_{i = 1}^p {{\beta _x}[i]X[n - i]} +\sum_{i = 1}^r {{\beta _z}[i]Z[n - i]} + u[n]\\
X[n] = \sum_{i = 1}^p  {{\beta _x}[i] X[n - i]} + \sum_{i = 1}^q {{\beta _y}[i]Y[n - i]}  \\\qquad+\sum_{i = 1}^r {{\beta _z}[i]Z[n - i]} + v[n].
\end{cases}\label{eqn:cGC1 and cGC2}
\end{equation} 
Then, the causal influence from $Y$ to $X$, conditional on $Z$, is defined as
\begin{equation}
{\mathcal{F}_{Y \to X|Z}} = \ln \frac{{{\mathop{\rm var}} (u)}}{{{\mathop{\rm var}} (v)}}.\label{F}
\end{equation}

Consider three model spaces $\mathcal{A}$, $\mathcal{B}$, and $\mathcal{C}$, where each space comprises three models. Let $p_i$ and $q_i$, $i=1,2,3$, denote the model orders regarding its endogenous information and the exogenous information from other variables, respectively. And $ n\in N, m\in N $,
\begin{align}
\mathcal{A}=\{ & a_1: p_1=n, q_1=0; \notag \\
& a_2: p_2=n+1, q_2=0; \notag \\
& a_3: p_3=n+2, q_3=0 \} \notag
\end{align}
\begin{align}
\mathcal{B}=\{ & b_1: p_1=n, q_1=m+0; \notag \\
& b_2: p_2=n, q_2=m+1; \notag \\
& b_3: p_3=n, q_3=m+2 \} \notag
\end{align}
\begin{align}
\mathcal{C}=\{ & c_1: p_1=n, q_1=m; \notag \\
& c_2: p_2=n+1, q_2=m; \notag \\
& c_3: p_3=n+1, q_3=m+1 \}. \notag
\end{align}
We further assume that $a_i$, $b_i$, and $c_i$, $i=1,2,3$, have the same residual variance. For space $\mathcal{A}$, the models only specify the regression model orders of the endogenous information by AIC/BIC, and then specify the causal effect of endogenous information by F-statistics. The models in space $\mathcal{B}$ specify the regression model orders of the exogenous effect by AIC/BIC, then specify the causal effect of exogenous information by F-statistics. But the models $\mathcal{C}$ specify the regression model orders of endogenous information and exogenous information separately, and specify the causal effect between $ c_1 $, $ c_2 $ and $ c_3 $ by F-statistics. In this situation, the model selection in conventional GCA is split into a two-stage scheme, the model orders is determining by AIC/BIC and then the causal effect is quantified by F-statistics sequentially. It is clear that the final inference might differ with rules applied, even though three classes are completely equivalent from a model description standpoint.

As stated above, specifying the effects of endogenous and exogenous information in conventional GCA is split into a two-stage scheme. Specifying the regression model orders of historical information, which contains the regression of endogenous information  in restricted model and the regression of both endogenous and exogenous information in unrestricted model, is mainly based on AIC/BIC theory separately, then specifying the causal effect of exogenous information by F-statistics. However, specifying the two effects is the same kind of problem of model selection from the mathematical perspective. Different theories might generated different benchmarks in two stages for model selection, therefore degrade the performance of GCA. 

Aside from theoretical considerations, there are still some issues in the F-test itself to be discussed. The American Statistical Association's (ASA) statement on statistical significance and $ p $-values has led to a collective rethinking of the entire scientific community, and many scientists believe that the application of $ p $-values in current scientific research has been distorted \cite{wasserstein2016asa,wasserstein2019moving,amrhein2019scientists,mcshane2019abandon}. This also leads many studies to selectively report results, and the dichotomous $ p $-value is arbitrary reductionism for scientific research. We believe that any reasonable study has its implied meaning, whether or not the $ p $-value is less than 0.05. We should understand the true meaning of $ p $-values, and it should not be over-exaggerated or degraded \cite{amrhein2018remove,amrhein2019inferential,hurlbert2019coup}. It is just a statistical tool, or just one of the mapping relationships that is purely mathematical. Specifically, pairwise $F$-statistics in conventional GCA arouses several potential problems that might lead to misleading or unreasonable inferences in connectivity analysis. Firstly, the model comparison with $F$-statistics is performed under a specific significance level. The assignment of significance level is a subjective matter in conventional GCA process \cite{miao2011altered,deshpande2011instantaneous}. A significance level that is too low could cause the false connection noise originated, whereas a significant level that is too high could erase the actual connection. When there is no rule to justify the assignment of the significance level, $F$-statistics will lead to different connectivity results depending on the significance level chosen.

\begin{figure*}
	\centering
	\subfloat[Search Path: $A \rightarrow C \rightarrow B$]{
		\includegraphics[width=4cm,height=3.5cm]{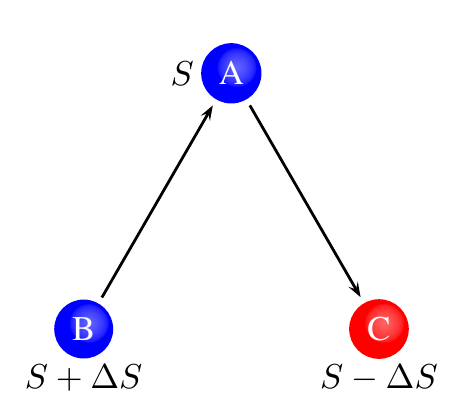}
	}
	\subfloat[Search Path: $B \rightarrow A \rightarrow C$]{
		\includegraphics[width=4cm,height=3.5cm]{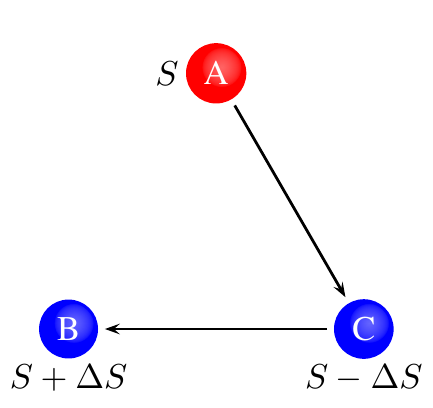}
	}
	\caption{Consider a conceivable case involving a collection of three candidate models $\mathcal{M}=\{A,B,C\}$ with residual variances of $S_A=S$, $S_B=S+\Delta S$, and $S_C=S-\Delta S$. Suppose $\frac{I}{2} \leq \Delta S < I$, where $I$ is the interval satisfying statistical significance. It is clear that model $C$, with the minimum variance, is the optimal model in this case. Following search path $B \rightarrow A \rightarrow C$ (a) we can arrive at optimal model $C$, while if we start with $A$ and follow the search path $A \rightarrow C \rightarrow B$ (b), we reach suboptimal model $A$. The results using $F$-statistics for model selection rely on both the search path and the initial model. Comparing transverses within the model space can only partially reduce the risk of model misspecification, and is not always available due to the nested relation between comparable models in $F$-statistics. Moreover, this strategy also leads to concern about the computational complexity.}
	\label{fig:search_path}
\end{figure*}

Secondly, the selection results by pairwise $F$-statistics are heavily dependent on the initially selected model and the search path in model space. Consider the collection of three models $\mathcal{M}=\{A,B,C\}$ and let $\mathcal{S}$ denote the residual variance of each model with $S_A=S$, $S_B=S+\Delta S$, and $S_C=S-\Delta S$. We further assume $\frac{I}{2} \leq \Delta S < I$, where $I$ is the interval satisfying statistical significance. The aim of $F$-statistics is to find the model appropriate to the observations from the model space, which is evidently model $C$ in this case. The search path $B \rightarrow A \rightarrow C$ will arrive at optimal model $C$, whereas if we start at $A$ and follow the search path $A \rightarrow C \rightarrow B$, we will reach suboptimal model $A$, that is, optimal model $C$ is not considered. The distinct results obtained by using $F$-statistics for model selection along with different search paths and initial models are given explicitly in Fig. \ref{fig:search_path}. In fact, $F$-statistics can not discriminate the models by residual variance within a specific range relative to the chosen significance level and the noise level. On the other hand, $F$-statistics uses the extra sum of squares principle to identify a better model. This means that only models with a nested relationship can be compared. Then the competitive models perform pairwise comparison in an indirect way, through $ intermediate $ model, namely unrestricted model in $F$-statistics. Therefore, the comparison between any two models is not available in practical applications, and the search path relies on the structure of all candidate models. In such situation the optimal model is not always guaranteed.

The third potential problem relates to computational complexity. Consider the simplest case where conditional causal connections are not taken into account ($q=1$ and $r=0$ in Eq. (\ref{eqn:cGC1 and cGC2})). Suppose that there are $m$ candidate models for any one directional causality in the network with $n$ nodes, the number of $F$-tests that needs to be performed is $m(m-1)(n-1)$, and the total number of comparisons is $mn(m-1)(n-1)$. The problem on computational complexity is compounded by conditional GCA, but it still will be intractable while investigating large network \cite{Zou_BMCB10}.

Although different approaches might be applied for model selection in GCA \cite{Kaminski,Li_TMI11}, the two stages are kept unchanged. They are generally based on two different mathematical theories in most GCA applications. However, as mentioned above, determining the model orders by AIC/BIC or quantifying the causal effects by F-statistics can essentially be considered as a generalized model selection problem from a modeling standpoint.  And model selection by a single mathematical principle throughout GCA, which could be easily ignored, would determine the final result working pattern of human brain. Thus the causal connectivity obtained by conventional GCA could be misleading, and main reason may be splitting model selection into a two-stage scheme. Since all the issues we discuss can be attributed to model selection problems, then a more practical model selection method need to be enabled.

To keep consistency in model selection for GCA, we proposed a code length guided framework based on MDL principle, which Rissanen first proposed to quantify parsimony of a model \cite{Rissanen1978Modeling}, to map the two-stage scheme into the same model space. Specifically, our proposal involves constructing a code length guided framework, then the model selection process in GCA can be converted into comparing the code length of each model. That means our proposal incorporated the endogenous and the exogenous information into a unified model selection process, which the information will be quantified by converting into code length to obtain causality. The two-stage scheme based on the two different theories is unified under the single mathematical framework, MDL principle, which guarantees the only benchmark in GCA methodology research.

And above all, the MDL is an information criterion that provides a generic solution for the model selection problem \cite{bryant2000model}. As a broad principle, the MDL represents a completely different approach for model selection relative to traditional statistical approaches, such as $F$-statistics and the AIC or BIC. Compared with the AIC/BIC, the use of the MDL does not require any assumptions about the data generation mechanism. In particular, a prior probability distribution does not need to be assigned to each model class. The objective of model selection in the MDL is not to estimate an assumed but unknown distribution, but to find models that more realistically represent the data \cite{Grunwald,Hansen_DECI99,Hansen2001Model}.

Fundamentally, MDL has intellectual roots in the algorithmic or descriptive complexity theory of Kolmogorov, Chaitin, and Solomonoff \cite{Li2008introduction}. Only considering the probability distribution as a basis for generating descriptions, Rissanen endowed MDL with a rich information-theoretic interpretation \cite{shannon1948mathematical,Rissanen1983A,rissanen1984universal,Rissanen1986Stochastic,Rissanen1987Stochastic,rissanen1996fisher}. Due to these characteristics, and it's reasonable to believe human brain meet minimum energy principle, causality analyzed with help of MDL principle may be more in line with physiological models of the brain.  On the whole,  our proposal takes the MDL principle as the single mathematical framework to select the generalized model for GCA, which to ensure the consistency, objectivity and the parsimony.

\section{The minimum description length principle}
\label{sec:the MDL principle}
The principle of \textit{parsimony} is the soul of model selection. To implement the \textit{parsimony} principle, one must quantify \textit{parsimony} of a model relative to the available data. With help of the work of Kolmogorov \cite{kolmogorov1965three,kolmogorov1968logical}, Wallace and Boulton \cite{Wallace1968An}, Rissenan formulated MDL as a broad principle governing statistical modeling in general. At beginning of modeling, all we have is only the data. Luckily, With the help of the algorithmic or description complexity theory of Kolmogorov, Chaitin, and Solomonoff, MDL regards a probability distribution as a descriptive standpoint. And MDL has some connections with AIC and BIC, sometimes it behaves similarly to AIC and BIC \cite{Rissanen,Hansen2001Model}. The difference is that MDL fixes attention on the length function rather than the code system. Therefore, many kinds of probability distribution can be compared in terms of their descriptive power \cite{Cover2012elements}. Our code length guided framework can be used in generalized model selection as long as there is a probability distribution in the model.

\subsection{Probability and idealized code length}

In order to describe the MDL principle explicitly, we deduced the formula of MDL in different cases. In model selection process of MDL, we need to compare the code length obtained by its probability distribution, so it is essential to understand the relationship between probability and the code length \cite{Hansen2001Model,grunwald2007minimum}. 

A code $ \varrho $ on a set $ \mathcal{A} $ is simply a mapping from $ \mathcal{A} $ to a set of codewords. Let $ \mathcal{A} $ be a finite binary set and let $ Q $ denote a probability distribution of the any element $ a $ in $ \mathcal{A} $. The code length of $ a $ is that $-\log_{2}Q$, the negative logarithm of $ Q $. For example, the Huffman code is one of the algorithms that constructed this relationship between probability and idealized code length \cite{Cover2012elements}.
Suppose that elements of $ \mathcal{A} $ are generated according a known distribution $ P $. Given a code $ \varrho $ on $ \mathcal{A} $ with length function $ L $, the expected code length of $ \varrho $ with respect to $ P $ is defined as
\begin{equation}L_{\varrho}=\sum_{x\subset A}P(x)L(x). \end{equation}
As is well known, if $ \varrho $ is prefix code, the code length $ L $ is equivalent to $ -\log_{2}Q(x) $ for some distribution $ Q $ on $ \mathcal{A} $. There is given an arbitrary code, if no codeword is the prefix of any other, the unique decodability is guaranteed. Any code satisfying this codeword condition is called a prefix code.

By Shannon's Source Coding Theorem, for any prefix code $ \varrho $ on $ \mathcal{A} $ with length function $ L $, the expected $ L_{\varrho} $ is bounded below by $ H(P) $, the entropy of $ P $. That is
\begin{equation} L_{\varrho} \geq H(P)=-\sum_{x\subset A}P(x)\log_{2}P(x), \end{equation}
where equality holds if and only if $L=-\log_{2}P$, in other words, the expected code length is minimized when $ Q=P $.

\subsection{Crude two-part code MDL}

In our view, modeling is a process that find the regularity of data and compress it. In model selection within MDL principle, What we have to do is selecting a suitable model based on the probability distribution of the object. Generally, the model we picked is overfitting or too simple. But model selection guided by the MDL principle, the complexity term or the error term in data fitting is incorporated into code length guided framework, which ensures the objectivity of the operation.

Until now, there are several forms of MDL principle to polynomial or other types of hypothesis and model selection. But at the original MDL, it usually divides the modeling for the data set into two parts, one part is to describe the model's self-information. The other is to describe the data set with the help of chosen probability model in part one. Consequently, here we firstly introduce the most common implementation of the idea -- the two part code version of MDL \cite{grunwald2007minimum,Hansen2001Model}.

Suppose the data $ D\in \mathcal{X}^{n} $ where $ \mathcal{X}=\{0,1\} $. Then there is a probability $ P\in \mathcal{M} $, and minimize
\begin{equation} L_{1,2}(P,D)=L_{1}(D|P)+L_{2}(P). \end{equation}
Here, it will select a reasonable model for $ D $ to make good predictions of future data coming from the same source, which therefore models the data using the class $ \mathcal{B} $ of all Markov chains \cite{grunwald2007minimum}.

\paragraph{The first part}
To get a better feel for the code $ L_{1} $, we prepare to consider two examples. First, let $ P_{\theta} \in B^{(1)} $ be some Bernoulli distribution with  $ P_{\theta}(x=1)=\theta $, and let $ D=(x_{1},\cdot\cdot\cdot,x_{N}) $. Since $ P_{\theta}(D)=\prod P_{\theta}(x_{i}) $ and $ \hat{\theta} $  is equal to the frequency of 1 in $ D $, the first part $ L_{1}(D|P) $ is given as
\begin{equation}\begin{aligned}
-\log P_{\theta}(D)&=-n_{1}\log\theta-n_{0}\log(1-\theta)\\&=-N[\hat{\theta}\log\theta+(1-\hat{\theta})\log(1-\theta)], \end{aligned}\end{equation} 
where $ n_{j} $ denotes the number of occurrences of symbol $ j $ in $ D $. Let $ k=2^{\gamma} $, the $ \gamma $th-order Markov chain model is denoted by $ \mathcal{B}^{(k)} $, it's defined as  $$ \mathcal{B}^{(k)}=\{P_{\theta}|\theta\in\varTheta^{(k)}\}; \varTheta^{(k)}=[0,1]^{k}. $$ 
Where $ \theta=(\eta[1|0\dots0],\eta[1|0\dots01],\dots,\eta[1|0\dots11]) $, for all $ n $, $ x^{n} $
\begin{equation} 
P_{\theta}(D)=(\frac{1}{2})^{\gamma}\prod_{i=\gamma+1}^{N}P_{\theta}(x_{i}|x_{i-1},\cdot\cdot\cdot,x_{i-\gamma}) \end{equation}
and
\begin{equation}\begin{aligned}-\log P_{\theta}(D)=-N\sum_{y\in\{0,1\}^{\log k}}(\hat{\eta}_{[1|y]}\log \eta[1|y]
\\+(1-\hat{\eta}_{[1|y]})\log (1-\eta_{[1|y]}))+\gamma.  \end{aligned}\end{equation}
Here $ \gamma=\log k $ is the number of bits needed to encode the first $ \gamma $ outcomes in $ D $. The maximum likelihood parameters $ \hat{\eta}_{[1|y]} $ are equal to the conditional frequencies of 1 prefixed by $ y $.
\paragraph{The second part}
In order to describe a $ P \in \mathcal{B} $, we have to describe a pair $ (k,\theta) $. We encode the all parameter in the distribution model by firstly encoding $ k $, which will use some prefix code $ C_{2a} $, and then code $ \theta $ with the help of the prefix code $ C_{2a} $. The resulting code $ C_{2} $(the code length of the first part) is then defined by $ C_{2}(k,\theta)=C_{2a}(k)C_{2b}(\theta|k) $. Firstly, $ \textbf{\textit{N}} $ is nature number set,
\begin{equation}  L_{2a}(k)=L_{\textbf{\textit{N}}}(k)=O(\log k). \end{equation}

Since $ \Theta^{(k)}=[0,1]^{k} $ is an uncountable infinite set, we would have to discretize it firstly. More precisely, we will restrict $ \Theta^{(k)} $ to some finite set $ \ddot{\Theta}^{(k)}_{d} $ of parameters that can be described using a finite precision of $ d $ bits per parameter.

Now that the number of elements of $ \ddot{\Theta}^{(k)}_{d} $ is  $ (2^{d})^{k} $. Thus it needs $ \log (2^{d})^{k}=k\cdot d $ bits to describe any particular $ \vartheta\in  \ddot{\Theta}^{(k)}_{d}  $, and may call $ d $ the precision used to encode a parameter. Letting $ d\vartheta $ be the smallest $ d $ such that $ \theta\in \ddot{\Theta}^{(k)}_{d} $, this gives
\begin{equation}
L_{2}(k,\theta)=
\begin{cases}
L_{\textbf{\textit{N}}}(k)+L_{\textbf{\textit{N}}}(d_{\vartheta})+kd_{\vartheta},  & \text{if $d_{\vartheta} <\infty $} \\
\infty, & \text{otherwise}.
\end{cases}\end{equation}

At the end, the crude two-part code MDL for Markov chain hypothesis selection is given by,
\begin{equation} \min_{k,d\in \textbf{\textit{N}};\theta\in \ddot{\Theta}^{(k)}_{d} }{-\log P_{k,d}(D)+kd+L_{\textbf{\textit{N}}}(k)+L_{\textbf{\textit{N}}}(d)}. \end{equation}

\section{Code length guided model selection in causality analysis}
\label{sec:MDL guided in GCA}
As stated above, conventional GCA splits the whole process into two stages which are actually modeling endogenous information and exogenous information. Since MDL has a close relationship with information theory, causality analysis with MDL here will also be more convincing and suitable. Further the regression of endogenous information can be converted to code length, and relative effect of exogenous variables can be also quantified by the code length guided framework, which the whole model selection process for GCA is unified into same model space.

The following is that MDL principle guided model selection in causality analysis\cite{Rissanen1978Modeling}, and variable $ {X_{N}} $ is given, 
\begin{equation} 
x_{t}=a_{1}x_{t-1}+a_{2}x_{t-2}+\cdot\cdot\cdot+a_{n}x_{t-n}+\epsilon_{t}.
\end{equation}
And the parameter vector consists of data
$ \theta=(n,\xi)$ and $\xi=(\sigma^{2},a_{1},\cdot\cdot\cdot,a_{n}) $, where  $ \sigma^{2}=\xi_{0} $ is the variance-parameter of zero-mean Gaussian distribution model for $ \epsilon_{t} $. And $ t=1,\cdot\cdot\cdot,m $, which $ m $ can be anyone more than $ n $ to keep the solution determined, $ N $ is the data length. In order to describe $ x_{t} $, turn to Gaussian distribution for $ \epsilon_{t} $. And $RSS=\sum_{t=1}^{m}(\epsilon_{t}^{2}) $ denotes the residual sum of squares corresponding to the estimation in the model. Clearly there is a Gaussian distribution model, applying the two part form of MDL, the total code length is given as 
\begin{equation}\begin{aligned}L(x,\theta)=m\ln\sqrt{2\pi} \sigma+\frac{RSS}{2\sigma^{2}}+\sum_{i=0}^{n}\ln \frac{|\xi_{i}|}{\delta}+\ln(n+1). 
\label{eqn:two-part MDL}
\end{aligned}\end{equation}
Where $ \delta $ is the precision, and it's optimal to choose $ 1/\sqrt{N} $ \cite{Rissanen1983A,Rissanen1989stochastic,Hansen2001Model}. Specially, $ \frac{|\xi_{i}|}{\delta}<1 $ should be ignored.

\subsection{Time-domain formulation}

Combining with the above formula, the causality investigation with the code length guided framework in the time domain can be carried out. There are two time-series ${X_{N}}$ and ${Y_{N}}$, then we consider two different models $ A $ and $ B $ (in Eq. (\ref{eqn:linear_restricted}) and Eq. (\ref{eqn:linear_unrestricted})) to describe $ {x_{t}} $. The representations are
\begin{equation}\begin{cases}
X_{t}=\sum_{j-1}^{n}a_{1i}X_{t-j}+\epsilon_{1t}\\
Y_{t}=\sum_{j-1}^{n}d_{1i}Y_{t-j}+\eta_{1t}\end{cases}
\label{eqn:linear_restricted}\end{equation}
where $ var(\epsilon_{1t})=\Sigma_{1} $ and $ var(\eta_{1t})=\Gamma_{1} $. Bivariate regressive representations are given, 
\begin{equation}\begin{cases}
X_{t}=\sum_{j-1}^{n}a_{2i}X_{t-j}+\sum_{j-1}^{n}b_{2i}Y_{t-j}+\epsilon_{2t}\\
Y_{t}=\sum_{j-1}^{n}c_{2i}X_{t-j}+\sum_{j-1}^{n}d_{2i}Y_{t-j}+\eta_{2t}
\end{cases}
\label{eqn:linear_unrestricted}\end{equation}
where $  var(\epsilon_{2t})=\Sigma_{2} $ and $ var(\eta_{2t})=\Gamma_{2} $, and their contemporaneous covariance matrix is 
\begin{equation*}
\boldsymbol\Sigma=\begin{pmatrix}
\Sigma_{2} & \Upsilon_{2} \\ 
\Upsilon_{2} & \Gamma_{2}
\end{pmatrix}\quad\text{where} \ \Upsilon_{2}=cov(\epsilon_{2t},\eta_{2t}).
\end{equation*}
Finally, $\epsilon_{1t}$ and  $\epsilon_{2t}$ have Gaussian distribution with mean 0 and unknown variance $ \sigma^{2} $, which denote the $ noise $ of time-series are fitting residual. Therefore, the distribution of residual terms $ \epsilon_{t} $ can be a standpoint to describe the model within MDL. Then, the code length of model $ A $ and $ B $ we obtained can be compared to identify the causal influence between $ x_{t} $ and $ y_{t} $. In the whole process of model selection for GCA, the causality investigation was mapped into the unified code length guided framework. According to Eq.(\ref{eqn:two-part MDL}), the code length of model $ A $ and model $ B $ can be given respectively. By the definition of Granger causality, the influence from $ Y $ to $ X $ is defined by our code length guided framework, 
\begin{equation}\begin{aligned}
F_{Y\rightarrow X}=L_{X}-L_{X+Y}.
\end{aligned}\end{equation}
Where $ L_{X}  $ denotes the code length of optimal model in Eq.(\ref{eqn:linear_restricted}), and $ L_{X+Y} $ denotes the code length of optimal model in Eq.(\ref{eqn:linear_unrestricted}). If $ F_{Y\rightarrow X} >0 $, it means causal influence from $ Y $ to $ X $ existed. Otherwise, there is no causal influence existed from $ Y $ to $ X $. As the causality represented above, our proposal can unify two-stage scheme into the code length guided framework, which can avoid inconsistency of two different mathematical theories or the subjectivity of F-statistics in model selection of the conventional GCA.

To compare conditional GCA, we consider the influence from $ Y $ to $ X $ while controlling for the effect from conditional node $ Z $ to $ X $. Firstly, the joint autoregressive representation is given
\begin{equation}\begin{cases}

X_{t}=\sum_{j-1}^{n}a_{3i}X_{t-j}+\sum_{j-1}^{n}b_{3i}Z_{t-j}+\epsilon_{3t}\\
X_{t}=\sum_{j-1}^{n}a_{4i}X_{t-j}+\sum_{j-1}^{n}b_{4i}Y_{t-j}+\sum_{j-1}^{n}c_{4i}Z_{t-j}+\epsilon_{4t}
\end{cases}\end{equation}
and $ var(\epsilon_{3t})=\Sigma_{3} $, $ var(\epsilon_{4t})=\Sigma_{3} $. 
By the definition of conditional GCA, if $ F_{Y\rightarrow X}>0 $ existed, causal influence from $ Y $ to $ X $ conditioned $ Z $ is defined
\begin{equation}\begin{aligned}
F_{Y\rightarrow X|Z}=L_{X+Z}-L_{X+Y+Z}.
\end{aligned}\end{equation}
Same as above, if $ F_{Z\rightarrow X}>0 $ existed, causal influence from $ Z $ to $ X $ conditioned Y is given
\begin{equation}\begin{aligned}
F_{Z\rightarrow X|Y}=L_{X+Y}-L_{X+Y+Z}
\end{aligned}\end{equation}

Clearly, in our code length guided framework, all candidate models can be compared freely in the context of their code length. Different from the traditional method, the causal connection is obtained by repeated pairwise comparison between models, our method can map all candidate models into the same model space without the repeating comparison and to obtain the conditional causal influence directly. Which is, if both $ F_{Y\rightarrow X}>0 $ and $ F_{Z\rightarrow X}>0 $ existed,
\begin{equation}\begin{aligned}
F_{Y,Z\rightarrow X}=min(L_{X+Y},L_{X+Z})-L_{X+Y+Z}.
\end{aligned}\end{equation}
Here if $ F_{Y,Z\rightarrow X}>0 $ existed, it means that both $ Y $ and $ Z $ have direct influence on $ X $. But if $ F_{Y,Z\rightarrow X} $is less than 0, there will be two cases. One is $ F_{Y,Z\rightarrow X}=(L_{X+Y}-L_{X+Y+Z})<0 $ existed, it means only $ Y $ has direct influence on $ X $. The other is $ F_{Y,Z\rightarrow X}=(L_{X+Z}-L_{X+Y+Z})<0 $ existed, it means that $ Z $ impacts $ X $ directly. In the unified model space, multiple selected models can be directly compared by code length, which can release the complexity of the algorithm. In this way, our proposal is more in line with Occam's razor, or the principle of \textit{parsimony}.

\subsection{Frequency-domain formulation}
With help of Geweke's work \cite{John1982Measurement}, the total interdependence between two time series $ X_{t} $ and $ Y_{t} $ can be decomposed into three components: two directional causal influences due to their interaction patterns, and the instantaneous influence due to factors possibly exogenous to the $ (X,Y) $ system (e.g. a common driving input) \cite{Guo2008Uncovering,Guo2010Granger}. Here other forms of regressive representations need to be considered, 
We first rewrite Eq.(\ref{eqn:linear_restricted}) and Eq.(\ref{eqn:linear_unrestricted})
\begin{equation}
\begin{pmatrix}
a_{2}(L) &b_{2}(L) \\ 
c_{2}(L) &d_{2}(L) 
\end{pmatrix}\begin{pmatrix}
X_{t}\\ 
Y_{t}
\end{pmatrix}=\begin{pmatrix}
\epsilon_{2t}\\ 
\eta_{2t}
\end{pmatrix}
\label{eqn:matrix_represent}
\end{equation}
where $ a_{2}(0)=1, b_{2}(0)=0, c_{2}(0)=0, d_{2}(0)=1 $, the lag operator $ L $ denotes $ LX_{t}=X_{t-1} $. Performing Fourier transform on both sides of Eq.(\ref{eqn:matrix_represent}), then we left-multiply
\begin{equation}
P=\begin{pmatrix}
1 & 0 \\ 
-\frac{\Upsilon_{2}}{\Sigma_{2}} & 1
\end{pmatrix}
\end{equation}
on both sides and rewrite the result equation, the normalized equations yield 
\begin{equation}
\begin{pmatrix}
X_{\omega}\\ 
Y_{\omega}
\end{pmatrix}=\begin{pmatrix}
D_{11}(\omega) &D_{12}(\omega) \\ 
D_{21}(\omega) &D_{22}(\omega) 
\end{pmatrix}\begin{pmatrix}
E_{2}(\omega)\\ 
H'_{2}(\omega)
\label{eqn:fre_normalize_matrix}
\end{pmatrix},
\end{equation}
where $ H'_{2}(\omega)=H_{2}(\omega)-\frac{\Upsilon_{2}}{\Sigma_{2}}E_{2}(\omega) $. The spectral matrix is $ \textbf{S}(\omega)=\textbf{D}(\omega)\boldsymbol\Sigma \textbf{D^{*}}(\omega) $, where $ * $ denotes complex conjugate and matrix transpose. The spectrum of $ X_{t} $ is 
\begin{equation}
S_{11}(\omega)=D_{11}(\omega)\Sigma_{2}D^{*}_{11}(\omega)-D_{12}(\omega)\Upsilon'_{2}D^{*}_{12}(\omega),
\label{eqn:spectrum_X}
\end{equation} 
where $ \Upsilon'_{2}=\Gamma_{2}-\frac{\Upsilon_{2}}{\Sigma_{2}}\Upsilon_{2} $. The first term in Eq.(\ref{eqn:spectrum_X}) is represented as the intrinsic influence and the second term as the causal influence of $ X_{t} $ due to $ Y_{t} $ at frequency $ \omega $. 
Based on this transformation, the causal influence from $ Y_{t} $ to $ X_{t} $ at frequency $ \omega $ is
\begin{equation}
f_{Y\longrightarrow X}(\omega)=ln\dfrac{S_{xx}(\omega)}{D_{11}(\omega)\Sigma_{2}D^{*}_{11}(\omega)}.
\label{eqn:fre_causality}
\end{equation}
The model orders of historical information in frequency domain for conventional GCA, are determined by AIC/BIC. Distinct from conventional GCA, we obtained the causal connectivities between nodes at frequency $ \omega $ by the code length guided framework.

\section{Experiments}
\label{sec:Experimental}
\subsection{$3$-node Network Simulation Experiments}

\subsubsection{Protocol for $3$-node Network}
To verify the performance of MDL principle in synthetic data experiment, a simple 3-node network is enabled. There were 1000 data points in time series of each node, to keep the stationarity of synthetic data, the first 700 data points were removed. The initial value of each node was 1, the variance of noise term $ \epsilon_i (i=1, 2, 3)$ varied from 0.15 to 0.35. The nodes was generated by   
\begin{equation}\begin{aligned}
\begin{cases}
y_{1,t}=1.5y_{1,t-1}-0.9y_{1,t-2}+\epsilon_{1}\\
y_{2,t}=0.8y_{1,t-1}+0.2y_{2,t-1}+\epsilon_{2}\\
y_{3,t}=-0.8y_{1,t-1}+0.4y_{3,t-1}+\epsilon_{3}
\end{cases}
\label{eqn:3-node}
\end{aligned}\end{equation}


\subsubsection{Results from $3$-node Network}
\begin{figure*}
	\centering
	\subfloat[]{
		\includegraphics[width=5cm,height=3.5cm]{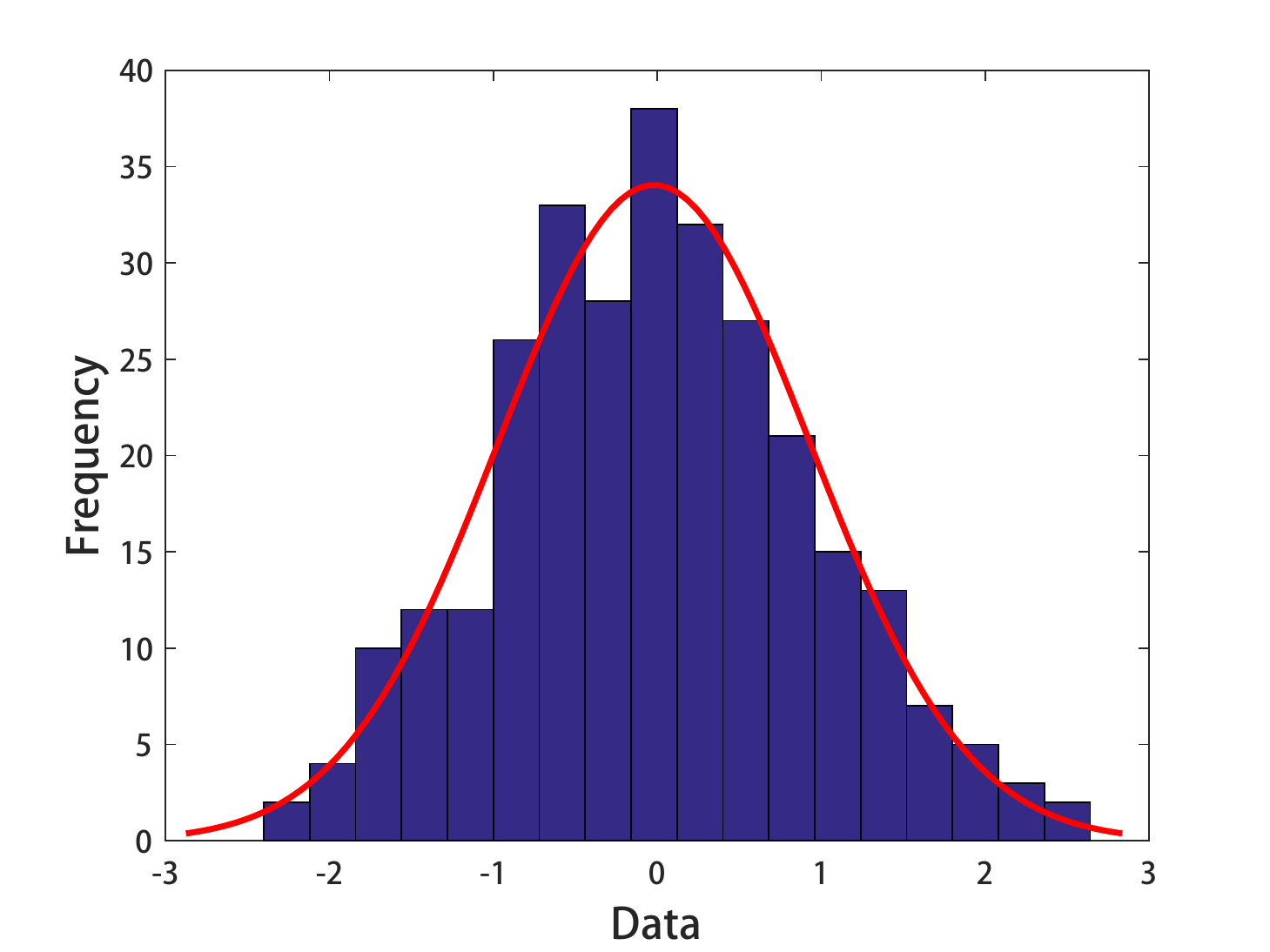}
		\label{fig:gaussian_fit}
	}
	\subfloat[]{
		\includegraphics[width=5cm,height=3.5cm]{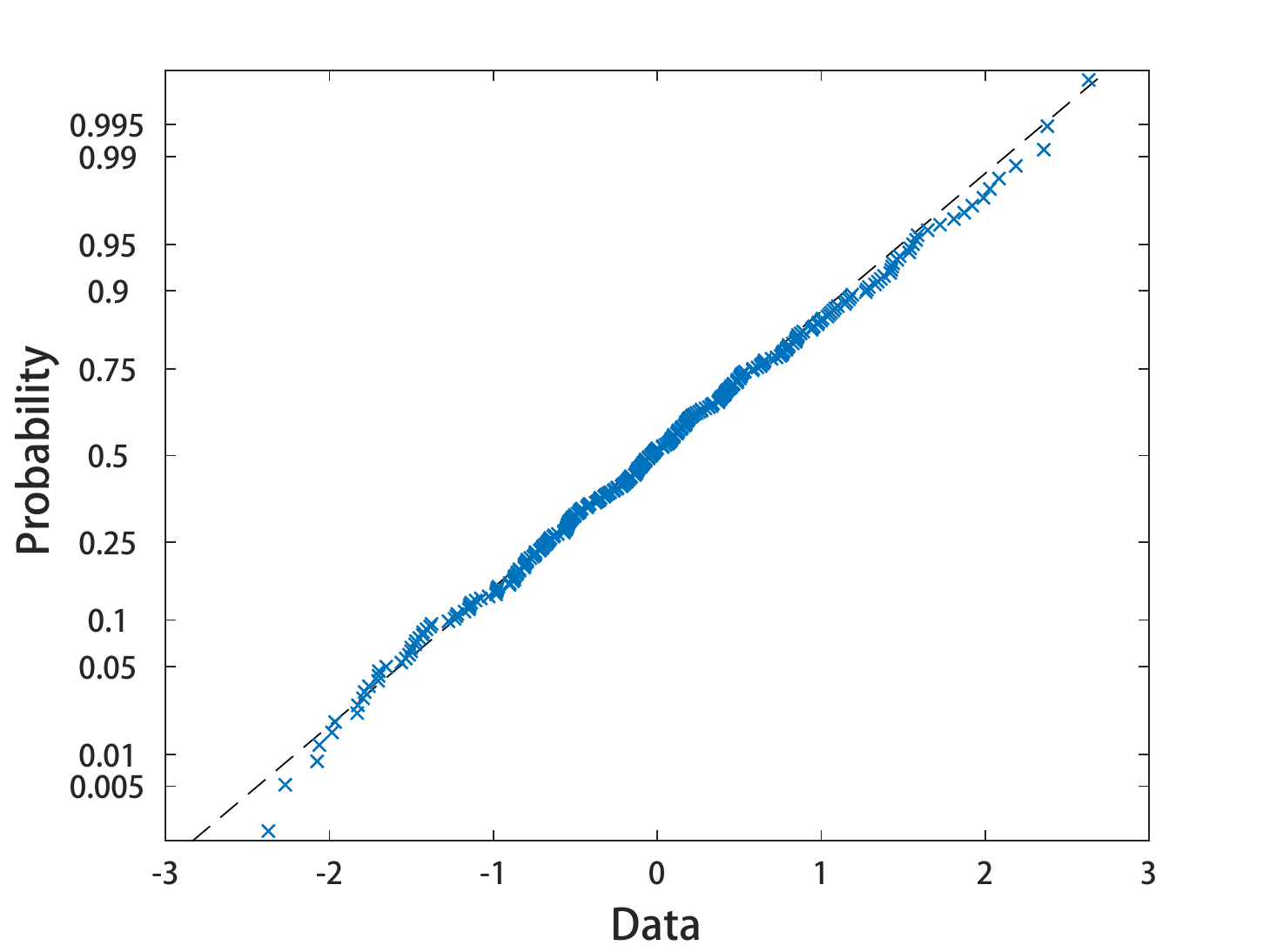}
		\label{fig:gaussian_histogram}
	}
	\subfloat[]{
		\includegraphics[width=5cm,height=3.5cm]{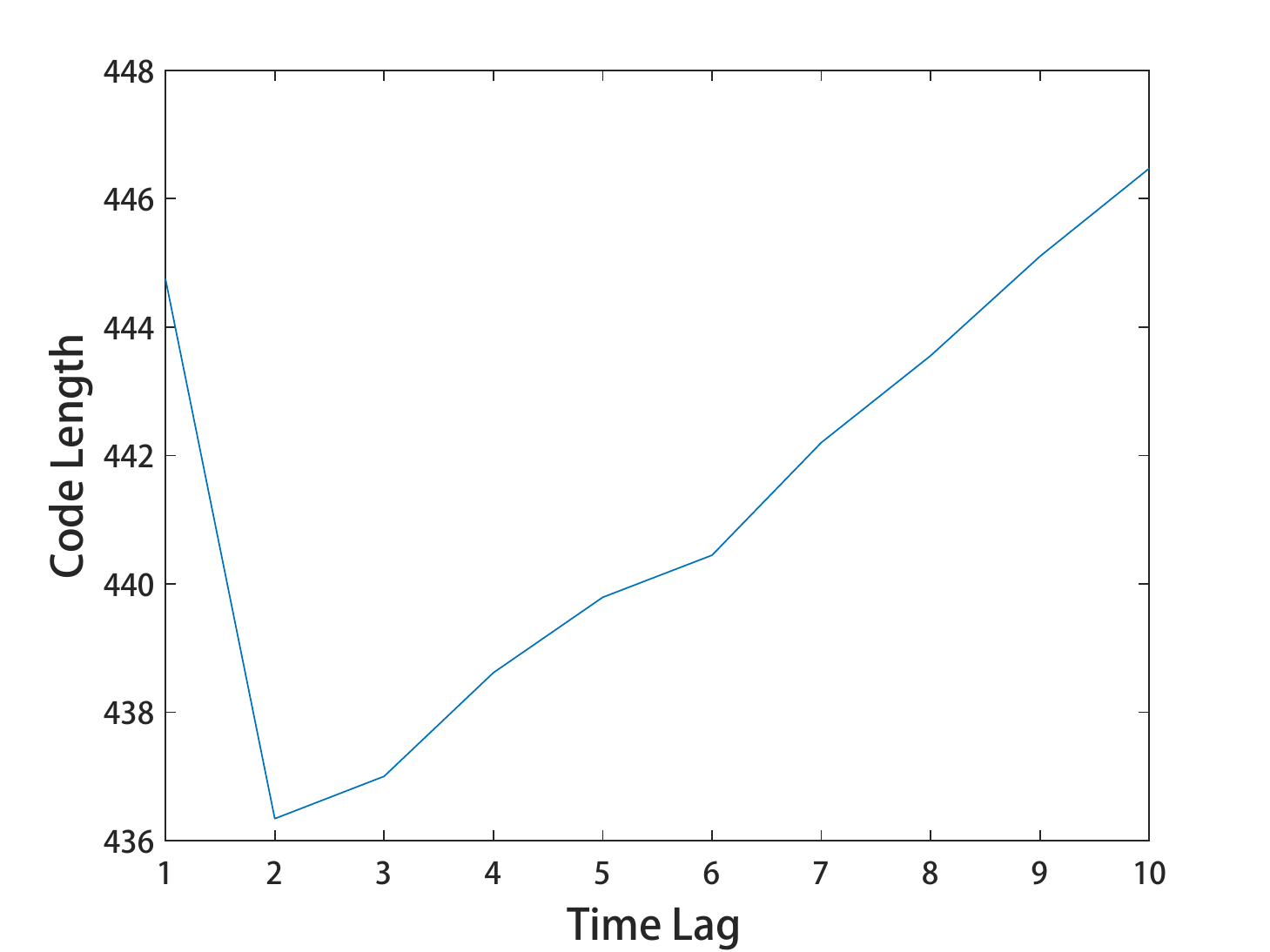}
		\label{fig:timelag}
	}
	\caption{The fitting Guassian distribution of the residual. (a): a histogram of values in data and fits a normal density function. (b): comparing the distribution of the data to the normal distribution. (c): code length obtained by our proposal, changed with different time lag in node 1.}
\end{figure*}

Firstly, to ensure the rigor of the proposal, the distribution of residual between the selected model and the observed data was verified. It was almost corresponding Gaussian distribution with mean 0, seen in Fig. \ref{fig:gaussian_fit} and \ref{fig:gaussian_histogram}, which guaranteed the validity for our initial description standpoint. And code length changed with different time lag in autoregressive model had shown in Fig. \ref{fig:timelag}, it dropped down to the minimum when time lag was 2, which corresponded with generated model in Eq.(\ref{eqn:3-node}).  
\begin{figure*}
	\small 
	\centering
	\subfloat[F-statistics $(P < 0.05)$]{
		\label{fig:3-node_results:GCA}
		\includegraphics[width=4cm,height=3.5cm]{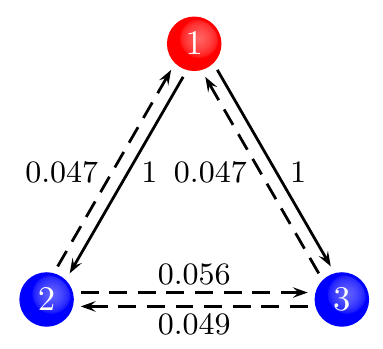}
	}	
	\subfloat[MDL method]{
		\label{fig:3-node_results:MDL}
		\includegraphics[width=4cm,height=3.5cm]{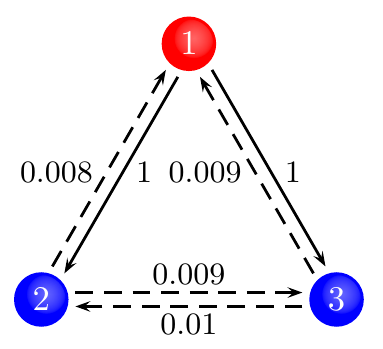}
		
	}
	\centering
	\caption{Comparison between two methods within low variance noise(0.15-0.35), the numbers on the arrow indicated the accuracy that the causal influence was identified in 1000 samples. Specially, the accuracy was only represent the probability that the single connection direction was identified by two methods.(a):causal network obtained by conventional GCA. (b):causal network obtained by our proposal.}
	\label{fig:3-node_results}
\end{figure*}
\begin{table*}[!ht]
	\footnotesize
	\centering
	\caption{Comparison between conventional GCA and MDL method in 3-node network}
	\begin{tabular}{ccccccc}
		\toprule
		\multirow{2}{*}{Noise level} & \multicolumn{2}{c}{\multirow{2}{*}{Method}} & \multicolumn{4}{c}{Accuracy(\%)}  \\
		\cmidrule(r){4-7}
		& & & Node 1      &  Node 2   &   Node 3	&  Total  \\
		\midrule
		\multirow{4}{*}{Low} &\multirow{3}{*}{F-statistics} & $ \alpha=0.1$  & 82.3 & 71.8 & 72.1 & 65.4 \\
		& & $ \alpha=0.05$  & 90.9 & 85.5 & 85.5 & 82 \\
		& & $ \alpha=0.01$  & 98.1 & 96   & 96.2 & 95.2 \\
		& \multicolumn{2}{c}{MDL}  & 98.4 & 97.3 & 97.2 & 96.5 \\
		\hline
		\multirow{4}{*}{Moderate} &\multirow{3}{*}{F-statistics} & $ \alpha=0.1$ & 81.4 & 72.5 & 71.8 & 64.8                 \\
		& & $ \alpha=0.05$  & 89.4 & 84.6 & 84.7 & 79.9 \\
		& & $ \alpha=0.01$  & 97.6 & 96.2 & 96.3 & 95.1 \\
		& \multicolumn{2}{c}{MDL}  & 98.5 & 97.7 & 97.4  & 96.8 \\
		\hline
		\multirow{4}{*}{High} &\multirow{3}{*}{F-statistics} & $ \alpha=0.1$ & 80.3 & 72  & 71.7 & 64.5                 \\
		& & $ \alpha=0.05$  & 89.5 & 83.6 & 83.7 & 79.1 \\
		& & $ \alpha=0.01$  & 98   & 96.7 & 96.7 & 95.7 \\
		& \multicolumn{2}{c}{MDL}  & 98.8 & 98   & 97.6 & 97.2 \\
		\bottomrule
		\multicolumn{7}{p{280pt}}{The variance of the low noise level data ranges from 1.5 to 3.5, and the moderate(high) level data ranges from 2.5 to 4.5(3.5-5.5). $ 1-\alpha $ denotes the significance level of F-test in conventional GCA. The accuracy of node i(i=1,2,3) only measures the causal connectivities with node i, not including the connections between the other two nodes. And the total accuracy represents the accuracy that the true model is found, it means only quantifies the causal influence from node 1 to node 2 and node 3.  }
		\label{tab:3-node_comparison}
	\end{tabular}
\end{table*}

Then, the causal connectivities identified by our proposal had shown in Fig. \ref{fig:3-node_results:MDL}, which obtained causal connection by comparing the code lengths according to Eq. (\ref{eqn:two-part MDL}), then the causal connectivities obtained by conventional GCA had shown in Fig. \ref{fig:3-node_results:GCA}. 
We found that causal influences from node 1 to node 2 and node 3 were identified both in GCA and our proposal. In other four causal influences, our proposal  almost guaranteed 99\% accuracy. But conventional GCA only guaranteed 95\% accuracy. 

\begin{figure*}
	\small 
	\centering
	\includegraphics[width=10cm,height=5cm]{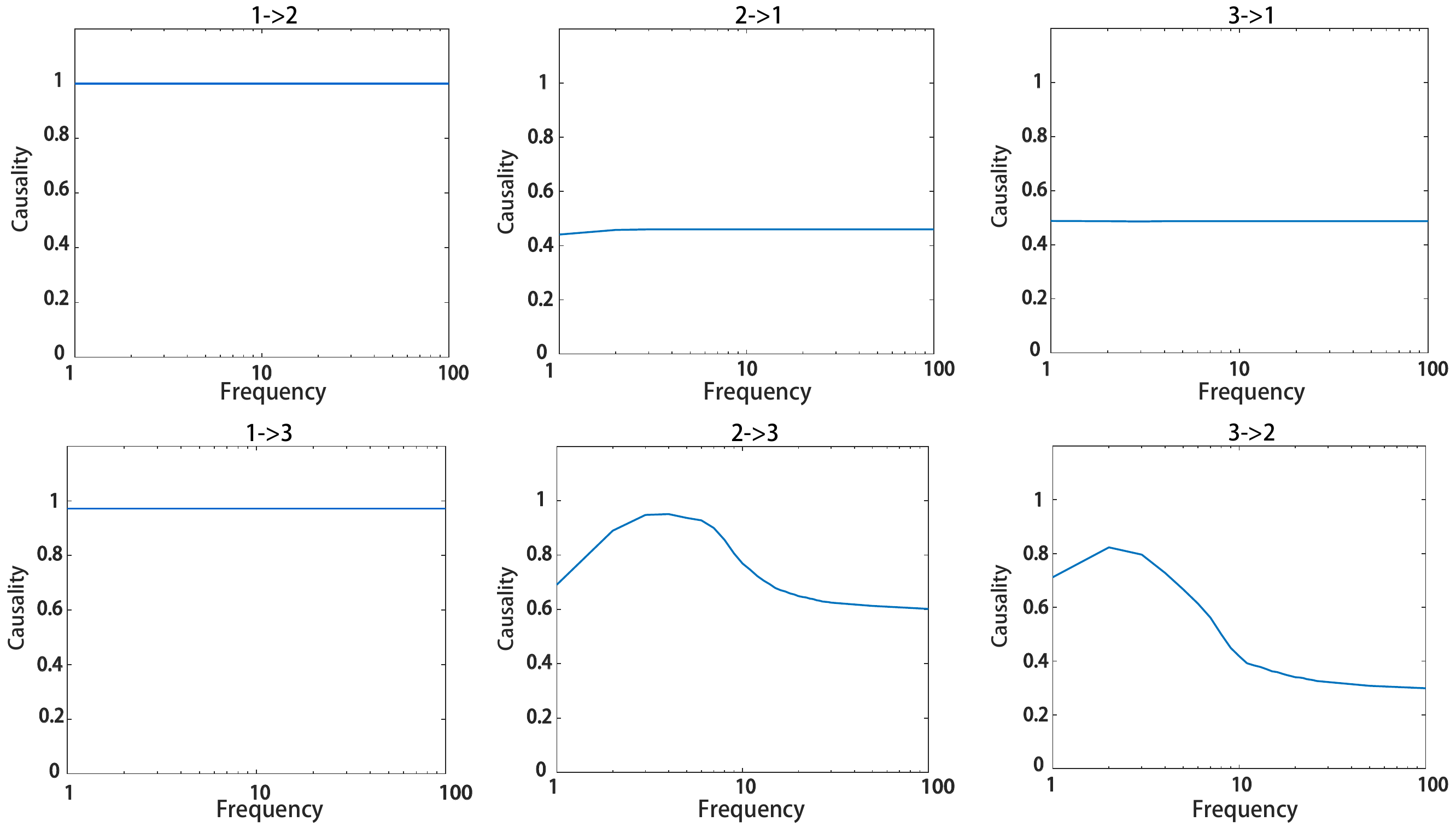}
	\caption{Causal connectivities between 3-node network obtained in frequency domain. Here we identified the causal influence at frequency $ \omega $ range from 1$ \sim $ 30, and at 50,100Hz.}
	\label{fig:fre_3-node}
\end{figure*} 
At the same time, experiments with three different noise levels and three confidence intervals had been carried out, the comparison between conventional GCA and our proposal showed in Table. \ref{tab:3-node_comparison}. The accuracy of causal connection associated with each node and the overall true model was also verified.
Comparing the accuracy of the single node and the overall network at the same noise level and the same confidence interval, it's unevenly distributed in conventional GCA. Meanwhile, the conventional GCA were very sensitive due to the different confidence intervals, especially in the identification of the overall network, even it had a stable performance at different noise level. In generally, our proposal always showed a relatively good performance whether it's identification in the whole network or a single node. It was worth noting that the results obtained by conventional GCA when $ \alpha=0.01 $ in F-statistics were close to the results of our proposal. The results were also in line with our expectations, that was because our proposal consider the complexity of the model more thoughtfully. As we emphasized above, there is only one mathematical principle guiding the model selection throughout the GCA process, thus our proposal is a more rigorous approach or a more robust approach. 

The oscillations in neurophysiological systems and neuroscience data are thought to constrain and organize neural activity within and between functional networks across a wide range of temporal and spatial scales. Geweke-Granger causality demonstrated that the oscillations at specific frequencies had been associated with the activation or inactivation of different encephalic region. But for conventional GCA in frequency domain, model selection of history information is determined by AIC or BIC. For our code length guided framework, the selected models for history information will regress into the $ true $ model space automatically. Due to its intellectual roots in descriptive complexity and close tie with information theory, our proposal may be more capable to identified causality in frequency domain. 

Same as time domain, causal connectivities at frequency $ \omega $ between 3-node network obtained by our proposal showed in Fig. \ref{fig:fre_3-node}.
In particular, causal connectivities in frequency domain were obtained within two nodes, which meant that we did not introduce conditional GCA in the frequency domain. This is mainly because there seems to be still some obfuscation with the method of using conditional GCA concept in the frequency domain. 
Therefore, as shown in Fig. \ref{fig:fre_3-node}, some non-existent connections between nodes were often misjudged, except for the causal influence $ 1\rightarrow2 $ and $ 1\rightarrow3 $. But we found that causal connectivities between node $2$ and node $3$ had a bigger chance to be misjudged at low frequency ($0-10$ Hz).	
Actually, comparing results obtained by conventional GCA in the time domain, causalities between two nodes were more legible in frequency domain, which meant only direct causalities showed more consistent results in our analysis. For example, there were only stable and significant causal influence existed in $ 1\rightarrow2 $ and $ 1\rightarrow3 $. 

\subsection{$5$-node Network Simulation Experiments}

\subsubsection{Protocol for $5$-node network }
To verified in more detail whether the proposed MDL method is just a conventional GCA method with a higher level of confidence, a complex 5-node network was given to further verified the robustness and the validity of our proposal, seen in Fig. \ref{fig:5-node}. Noise terms $ \epsilon_i (i=1, 2,..., 5) $ were the Guassian distribution with mean 0, and the variance ranged from 0.15 to 0.3. The first two initial values of nodes are 1, they were given by
\begin{equation}\begin{aligned}
\begin{cases}
x_{1,t}=0.792x_{1,t-1}-0.278x_{1,t-2}+\epsilon_{1}\\
x_{2,t}=0.768x_{2,t-1}-0.503x_{2,t-2}+0.83x_{1,t-1}\\-0.32x_{1,t-2}+\epsilon_{2}\\
x_{3,t}=0.67x_{3,t-1}-0.312x_{3,t-2}+0.56x_{2,t-1}\\-0.42x_{2,t-2}+\epsilon_{3}\\
x_{4,t}=0.733x_{4,t-1}-0.27x_{4,t-2}+0.72x_{2,t-1}\\-0.27x_{2,t-2}+0.52x_{3,t-1}-0,456x_{3,t-2}\\\qquad+0.76x_{5,t-1}-0.33x_{5,t-2}+\epsilon_{4}\\
x_{5,t}=0.845x_{5,t-1}-0.24x_{5,t-2}+0.68x_{4,t-1}\\-0.254x_{4,t-2}+\epsilon_{5}
\label{eqn:5-node}
\end{cases}
\end{aligned}\end{equation} 

\subsubsection{Results from $5$-node Network}	

\begin{figure*}[!ht]
	\small 
	\centering
	\includegraphics[width=7cm,height=3.5cm]{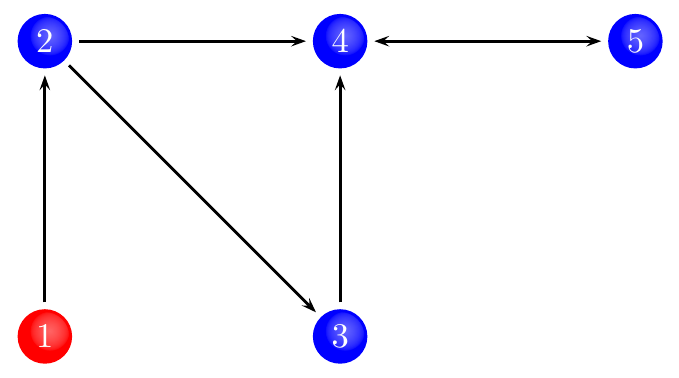}
	\caption{The relationships of simulation data sets in 5-node network.}
	\label{fig:5-node}
\end{figure*} 
\begin{figure*}[!ht]
	\centering
	\subfloat[F-statistics ($ P<0.05 $)]{
		\includegraphics[width=7.5cm,height=5cm]{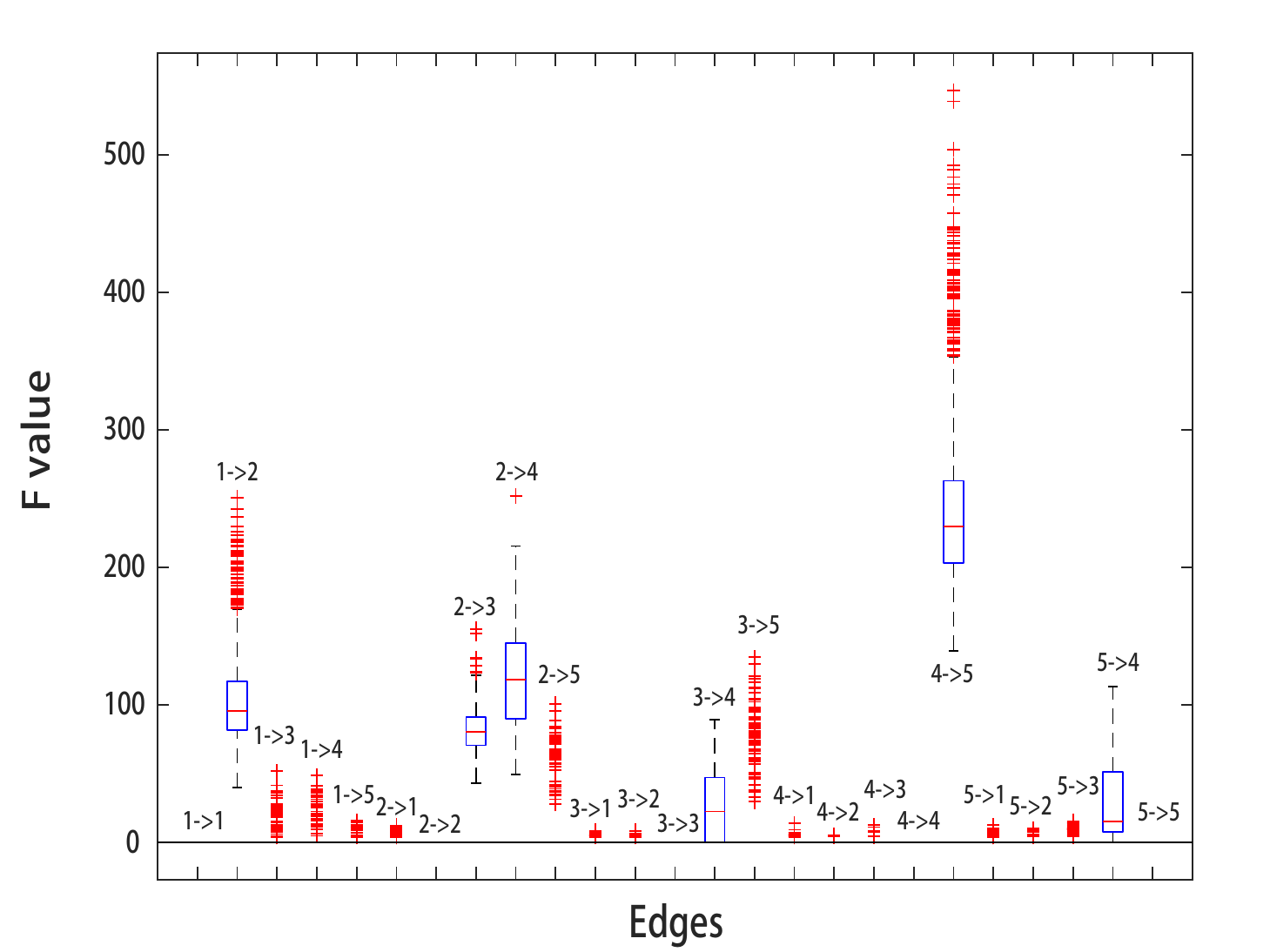}
		\label{fig:5-node_cGCA}
	}
	\subfloat[MDL method]{
		\includegraphics[width=7.5cm,height=5cm]{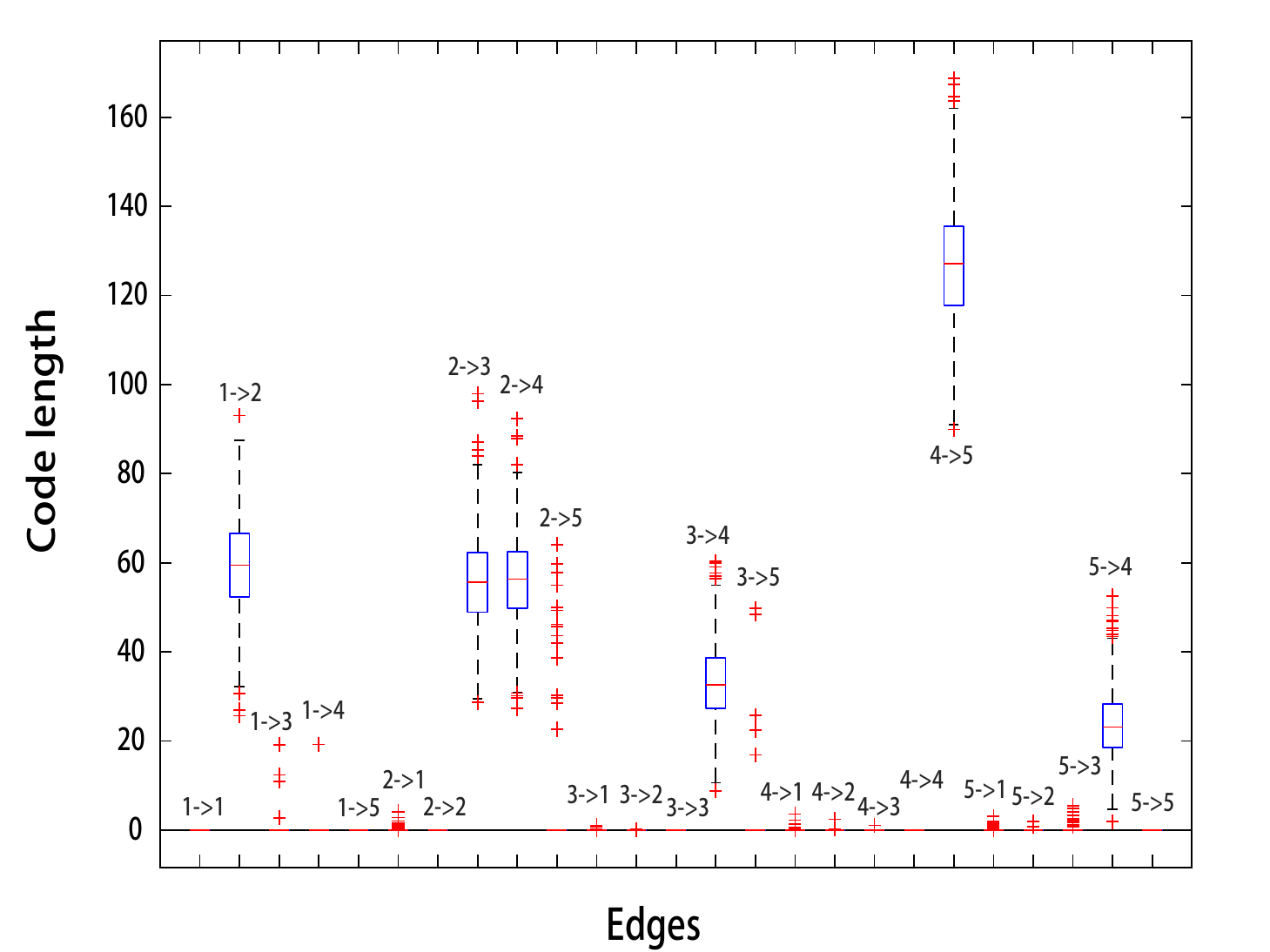}
		\label{fig:5-node_MDL}
	}
	\caption{Causal connectivities between 5 nodes identified by conventional GCA and our proposal respectively. (a): causal connection network was obtained by conventional GCA at $ \alpha=0.05 $ in F-test. (b): causal connection network obtained by our proposal.}
	\label{fig:5-node_results}
\end{figure*}
\begin{table*}[!ht]
	\footnotesize 
	\centering
	\caption{ 5-NODE NETWORKS}
	\label{table}
	\setlength{\tabcolsep}{5pt}
	\begin{tabular}{p{60pt} p{40pt}p{80pt} p{45pt} p{50pt}}
		\hline
		\multicolumn{2}{c}{Edge}&F-statistics$ (\alpha =0.05)$ &$ ( \alpha =0.01)$& MDL method \\
		\hline
		\multirow{6}{*}{true connections}&$ 1\longrightarrow2 $&1000/1000    & (1000/1000)& 1000/1000\\
		&$ 2\longrightarrow3 $&1000/1000& (1000/1000)& 1000/1000\\
		&$ 2\longrightarrow4 $&1000/1000& (1000/1000)& 1000/1000\\
		&$ 3\longrightarrow4 $&507/1000 & (456/1000) &1000/1000\\
		&$ 4\longrightarrow5 $&1000/1000 & (1000/1000)   & 1000/1000\\
		&$ 5\longrightarrow4 $&929/1000 & (780/1000) & 1000/1000\\
		\hline
		\multirow{14}{*}{false connections }
		&$ 1\longrightarrow3 $&36/1000& (8/1000)   &  4/1000\\
		&$ 1\longrightarrow4 $&43/1000 & (1/1000)   &  1/1000\\
		&$ 1\longrightarrow5 $&11/1000 & (0/1000)   &  0/1000\\
		&$ 2\longrightarrow1 $&54/1000 & (12/1000)  & 15/1000\\
		&$ 2\longrightarrow5 $&58/1000 & (14/1000)  & 15/1000\\
		&$ 3\longrightarrow1 $&44/1000 & (17/1000)  & 7/1000\\
		&$ 3\longrightarrow2 $&15/1000 & (2/1000)   & 2/1000\\
		&$ 3\longrightarrow5 $&83/1000 & (29/1000)  &5/1000\\
		&$ 4\longrightarrow1 $&30/1000 & (8/1000)   & 14/1000\\
		&$ 4\longrightarrow2 $&6/1000 & (0/1000)& 2/1000\\
		&$ 4\longrightarrow3 $&5/1000 & (1/1000)& 1/1000\\
		&$ 5\longrightarrow1 $&53/1000 & (18/1000)  & 16/1000\\
		&$ 5\longrightarrow2 $&17/1000 &(1/1000)   & 2/1000\\
		&$ 5\longrightarrow3 $&41/1000 & (8/1000)   & 11/1000\\
		\hline
		\multicolumn{5}{p{300pt}}{Causal connectivities between nodes in 5-node network within low variance noise(0.15-0.3). True connection represents the causal influence truly existed in Fig. \ref{fig:5-node}, false connection means not existed.}	\end{tabular}
	\label{tab:5-node_edges}  
\end{table*}
The result of connections between 5 nodes showed in Fig. \ref{fig:5-node_MDL}, causal influence analyzed by code length was largely consistent with connections in Fig. \ref{fig:5-node}. Simultaneously, causalities analyzed by conventional GCA between 5 nodes showed in Fig. \ref{fig:5-node_cGCA}. Same as results in 3-node network, our proposal showed 100\% consistency with connections in Fig. \ref{fig:5-node} between directly causal related nodes, seen in Table. \ref{tab:5-node_edges}. 
\begin{figure*}
	\small 
	\centering
	\includegraphics[width=16cm,height=12cm]{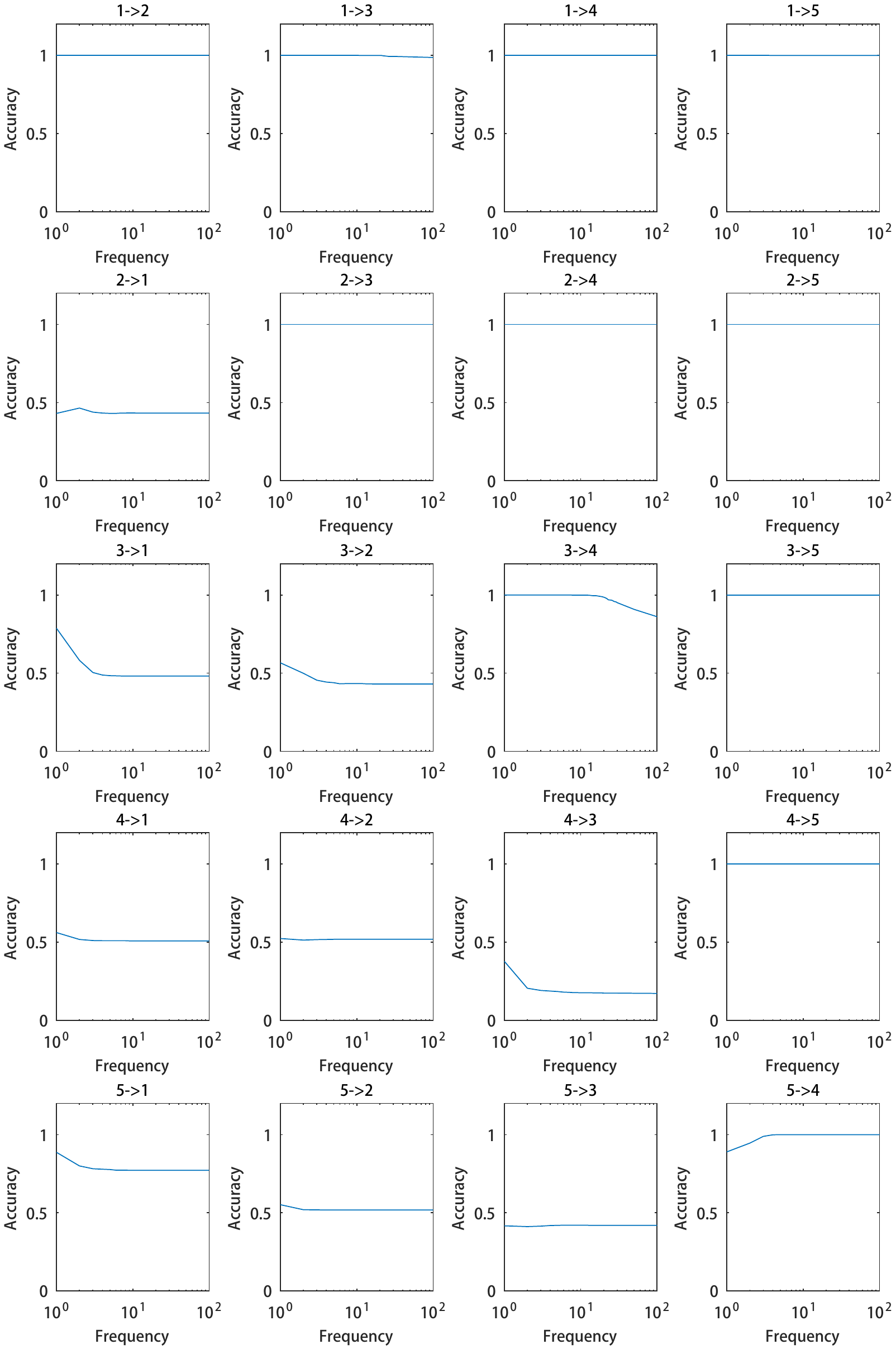}
	\caption{Causal connectivities of 5-node network obtained in frequency domain. Same as 3-node network, we identified the causal influence at frequency $ \omega $ range from 1$ \sim $ 30, and at 50, 100Hz. }
	\label{fig:fre_5-node}
\end{figure*}
And in other connections in 5-node network, the accuracy of our proposal also was not below 98.4\%. Whereas, conventional GCA did not showed the same robustness of our proposal. Causal connectivities between direct related nodes was not well identified, for example only 507 samples were identified as causal influences from node 3 to node 4 in 1000 samples and causal influence from node 5 to node 3 was identified at 92.9\% accuracy. 
And in other connections, the accuracy of conventional GCA was more poor than our proposal. Clearly, in higher confidence levels ($ \alpha=0.01 $) of conventional GCA, although the specificity of causal identification increased, its sensitivity decreased. 

More importantly, whether the causal influence from Node 3 to Node 4 or from Node 5 to Node 4, the significance level of connection were not enough to be identified, even for $ \alpha=0.01 $ in F-statistics. Therefore, the results demonstrated that our proposal was not equivalent to the conventional GCA with a higher significance level in F-statistics at all. Obviously, when target network was more complicated in simulation, our proposal showed a more desirable property in time domain, while conventional GCA generally made mistakes. Luckily, our proposal performed very well regardless of the existence fo relationships between nodes, even as in more complicated network. Same as 3-node network, the result causal network was unrelated with the varying variance of noise from (0.15-0.3) to (0.35-0.5). In conditional causality analysis, our approach reduced the complexity of algorithm, eliminating the need for repeated pairwise comparisons between models, while ensured the accuracy of results.

Subsequently, we identified the causal connectivities among 5-node network in frequency domain by our proposal, seen in Fig. \ref{fig:fre_5-node}. Same as the 3-node network, causal connection networks were identified without introducing conditional GCA, and the causal connection network had regular characteristics. The causal influence whether it was direct or indirect existed in 5-node network was more stable to be identified in the frequency domain. Similarly, since conditional GCA was not considered, other non-existent causal connectivities had a chance to be identified. And in 5-node network, the possibility of being misjudged was even greater. Therefore, it is necessary to introduce conditional GCA to distinguish direct from indirect influences between system components in the frequency domain. And at the same time, we found that removing the noise frequency component is an obstacle to causality analysis, main reason is that we have no prior knowledge about which one is \textit{noise} or others in row data.

\begin{figure*}
	\small 
	\centering
	\subfloat[visual (CSA-control)]{
		\begin{minipage}[t]{0.5\linewidth}	
			\centering
			\includegraphics[scale=0.16]{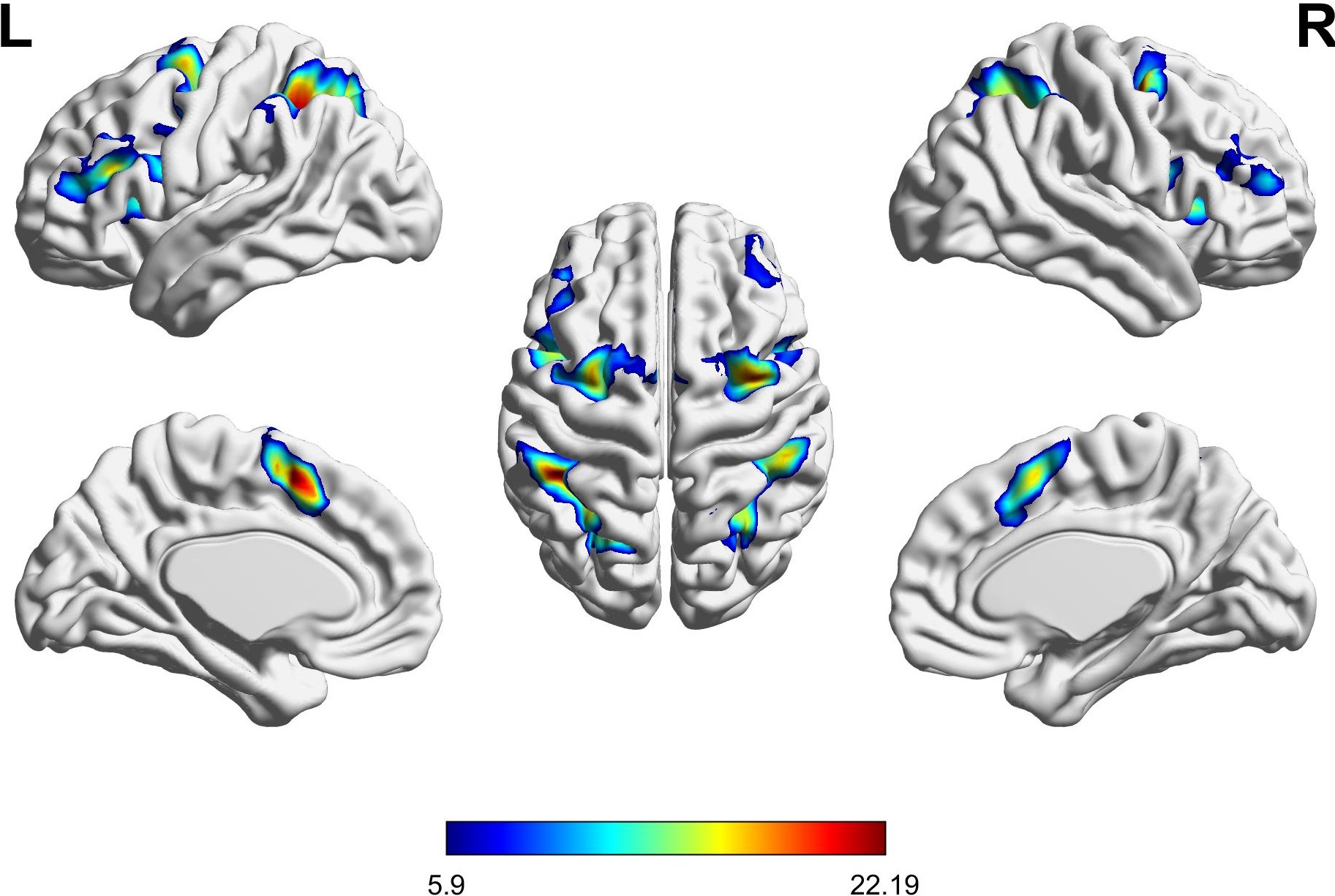}
		\end{minipage}%
	}	
	\subfloat[auditory (CSA-control)]{
		\begin{minipage}[t]{0.5\linewidth}	
			\centering
			\includegraphics[scale=0.16]{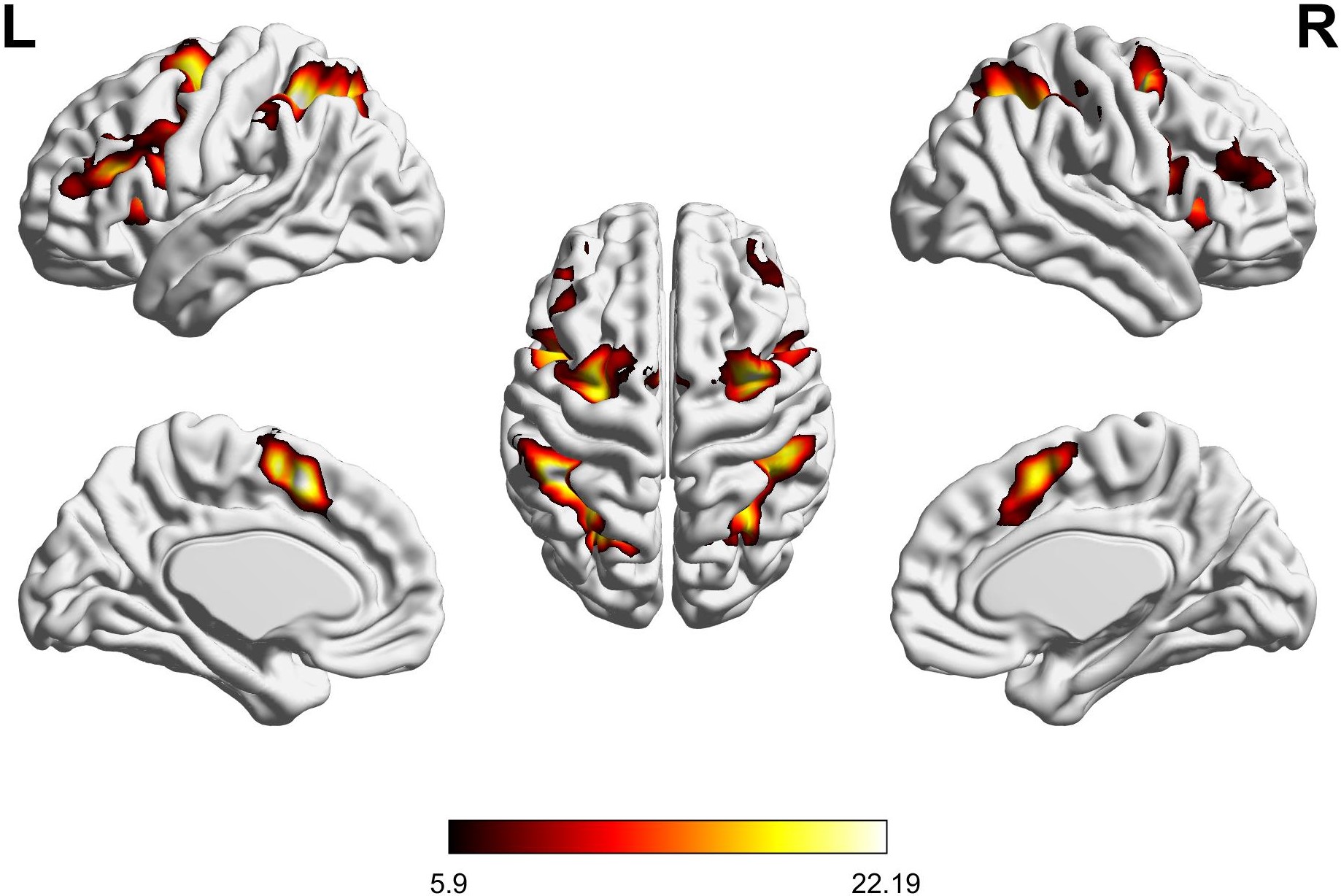}
		\end{minipage}%
	}
	\centering
	\caption{Mental calculation of CSA-control state under the two stimuli(visual stimulus and auditory stimulus), the activation regions were processed by SPM12, the control state meant that the sample was in rest state without mental calculation. (a): CSA-control state under visual stimulus. (b): CSA-control state under auditory stimulus.(P$ < $0.0001, uncorrected)}
	\label{fig:activation regions_mental calculation}
\end{figure*}
\begin{figure*}[!ht]
	\small 
	\centering
	\includegraphics[width=12cm,height=8cm]{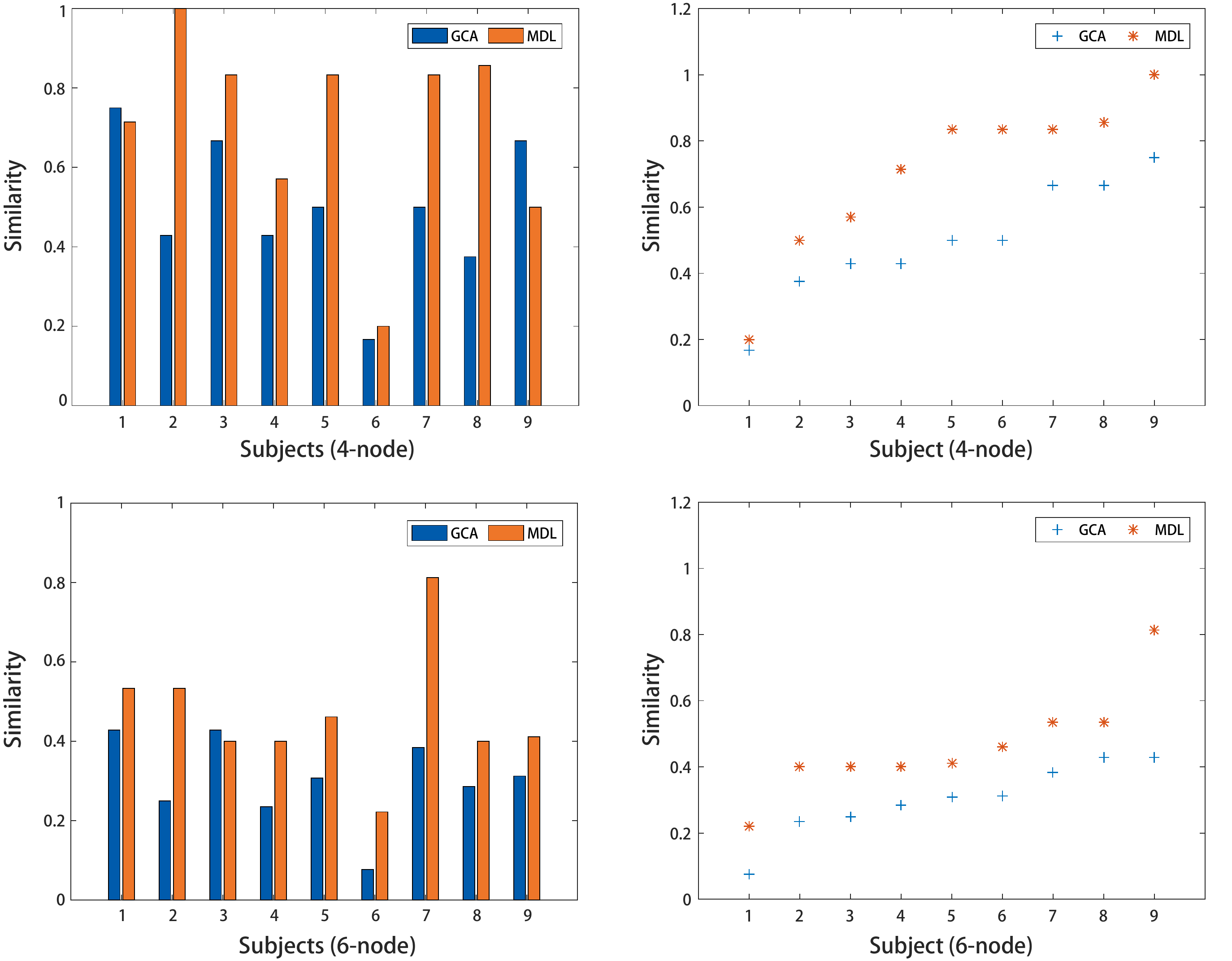}
	\caption{The similarity in 4-node mental calculation network and 6-node full network respectively. Left panel compared the similarity on the individual, and the right panel compared the distribution of similarities, collectively.}
	\label{fig:similarity}
\end{figure*}
\subsection{Real data}
\subsubsection{Experimental protocol }
In the study we let ten subjects perform simple one-digit(consisting of 1-10) serial addition (SSA) and complex two-digit(consisting of 1-5) serial addition (CSA) by visual stimulus and simultaneously measured their brain activities with fMRI.

Immediately, we asked the subjects to perform same serial addition arithmetic tasks by auditory stimulus. Nine right-handed healthy subjects (four female, 24$ \pm $1.5 years old) and one left-handed healthy female subject (24 years old) participated. One of the subject's(a right hand male) data was deleted due to excessive head motion. All subjects volunteered to participate in this study with informal written consent by themselves. 

\begin{figure*}
	\small 
	\centering
	\subfloat[Statistics (visual)]{
		\begin{minipage}[t]{0.4\linewidth}	
			\centering
			\includegraphics[scale=0.45]{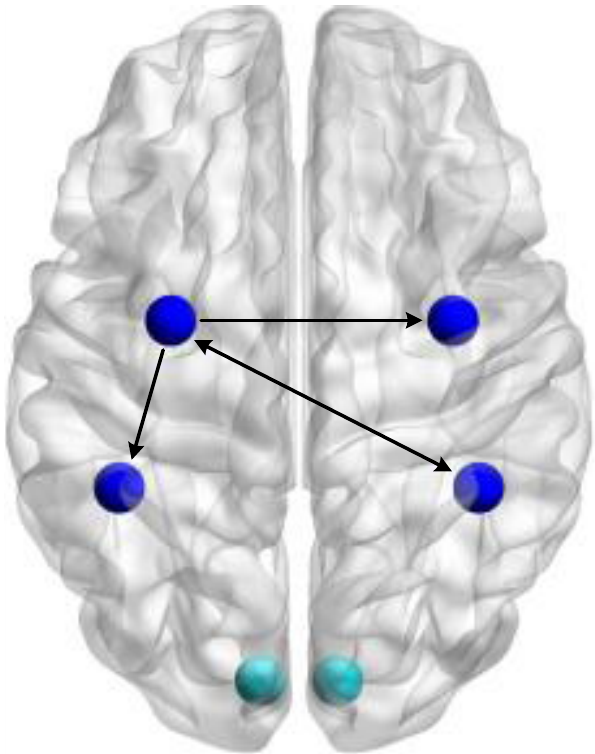}
		\end{minipage}%
	}
	\small 
	\centering
	\subfloat[Statistics (auditory)]{
		\begin{minipage}[t]{0.5\linewidth}	
			\centering
			\includegraphics[scale=0.45]{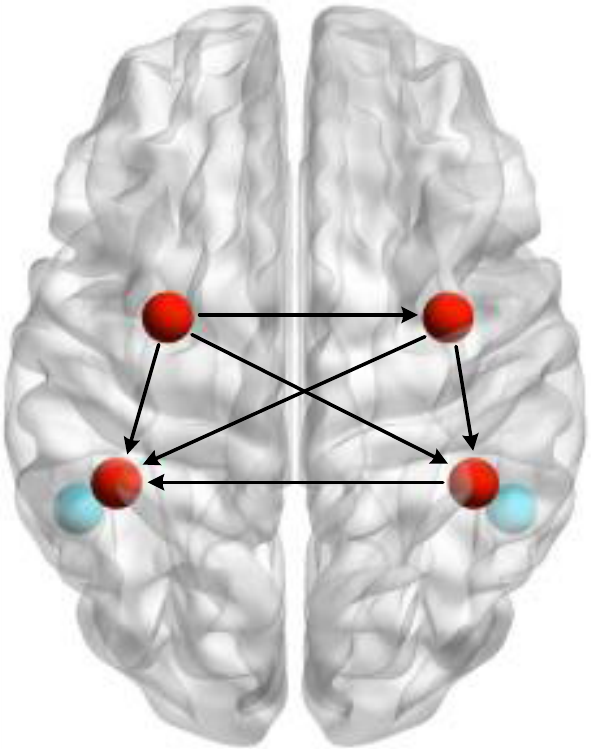}
		\end{minipage}%
	}
	
	\subfloat[MDL method (visual)]{
		\begin{minipage}[t]{0.4\linewidth}	
			\centering
			\includegraphics[scale=0.45]{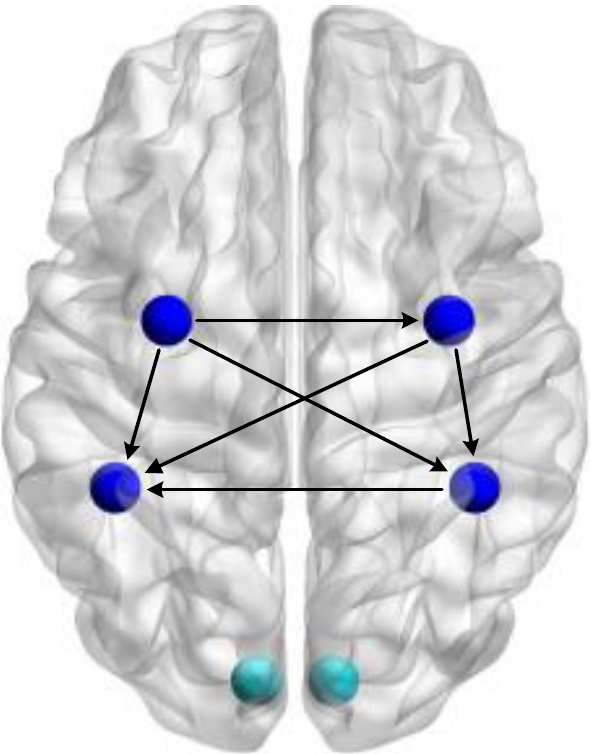}
		\end{minipage}%
	}
	\small 
	\centering
	\subfloat[MDL method (auditory)]{
		\begin{minipage}[t]{0.5\linewidth}	
			\centering
			\includegraphics[scale=0.45]{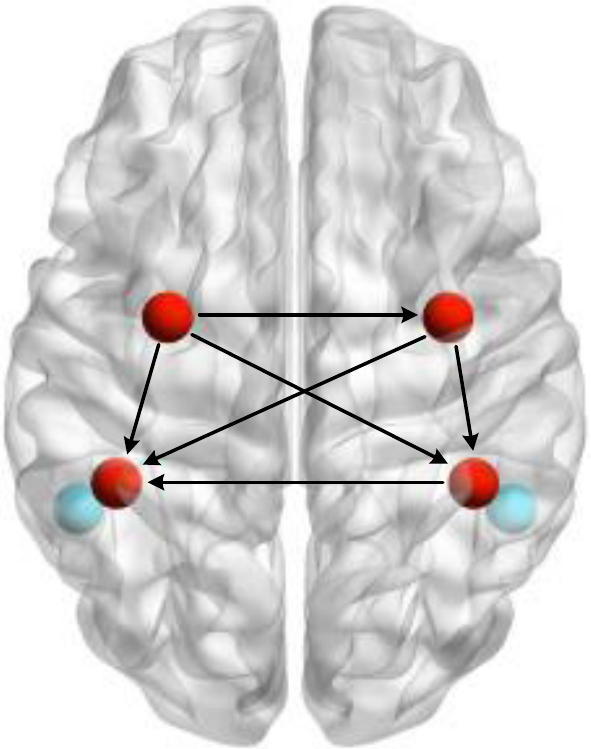}
		\end{minipage}%
	}	
	\centering
	\caption{Causal connectivities of subject 2 in mental calculation network under two stimuli. Cyan node represented visual/auditory stimulus node. The blue/red nodes related to the mental calculation network stated in the text. (a) and (b): mental calculation network obtained by conventional GCA. (c) and (d): mental calculation network obtained by proposed approach.}
	\label{fig:subject2_mental calculation network}
\end{figure*}
The causal connection network of one subject performing same task should be same or similar, which is also the basic assumption in group analysis. Therefore, for mental calculation under different stimuli, causal connection network of same subject may be different in the input node of stimulus, but we have reason to postulate that the causal connection should be same inside the mental calculation network, at least it should be similar. Thus, we will compare the similarity of causal connection network within the same mental calculation task under different stimuli. More directly, in order to quantify which method is more robust, we will compare the similarity of the result network within auditory and visual stimulus on each subject. To obtain the similarity between networks, we define a measure method to quantify the similarity, 
\begin{equation}
S=
\dfrac{\sum\sum(A\cap B)}{\sum\sum(A\cup B)}.
\end{equation}
Where $ S $ represents the measured similarity, $ A $ and $ B $ are the connected matrix of mental calculation networks. The numerator is the sum of intersections of the same connected edges, and the denominator is the sum of the unions of the connected edges.

\subsubsection{Mental calculation network}
We already have verified the validity of MDL in simulation data, but behavior in real data will determine the truly robustness of the methods. Firstly, the activation regions of mental calculation under different stimuli showed in Fig. \ref{fig:activation regions_mental calculation}, and as had been postulated above, the activation regions between two senses were identical. The similarities in 4-node connection network and 6-node full connection network showed in Fig. \ref{fig:similarity} respectively. In 4-node network, we found that causal connection networks of mental calculation between visual stimulus and auditory stimulus were very similar for most subjects, except for subject 6. There were seven of nine subjects the similarities were above 0.6. Turned to conventional GCA, we found that only three subjects of causal connection networks between the two stimuli had a similarity above 0.6. Meanwhile, only in subject 1 and subject 9, we found the similarity of causal networks between two senses was above our proposal. Even for these two subjects, the similarities in our proposal were close to conventional GCA, especially in subject 1.
Further the similarities in 6-node full connection network also showed in Fig.11. Duo to the difference between the input node of stimulus, the similarities in two methods were not above 0.5 mostly, but there were still 3 subjects the similarities in our proposal were mostly above 0.5, especially for subject 7. Clearly, whether inside connection network of mental calculation or among the 6-node full connection network, the similarities in most of our subjects identified by our proposal were more than conventional GCA.

\begin{figure*}
	\small 
	\centering
	\subfloat[Statistics (visual)]{
		\begin{minipage}[t]{0.4\linewidth}	
			\centering
			\includegraphics[scale=0.45]{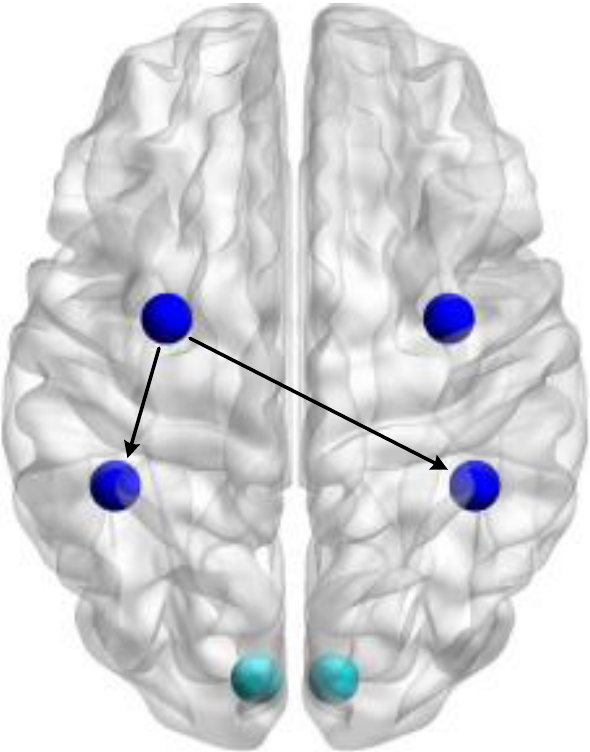}
		\end{minipage}%
	}
	\small 
	\centering
	\subfloat[Statistics (auditory)]{
		\begin{minipage}[t]{0.5\linewidth}	
			\centering
			\includegraphics[scale=0.45]{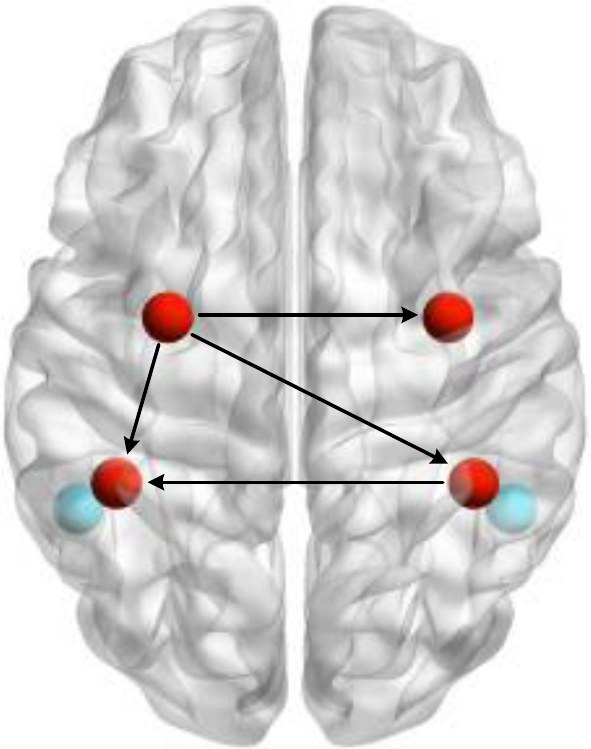}
		\end{minipage}%
	}
	
	\subfloat[MDL method (visual)]{
		\begin{minipage}[t]{0.4\linewidth}	
			\centering
			\includegraphics[scale=0.45]{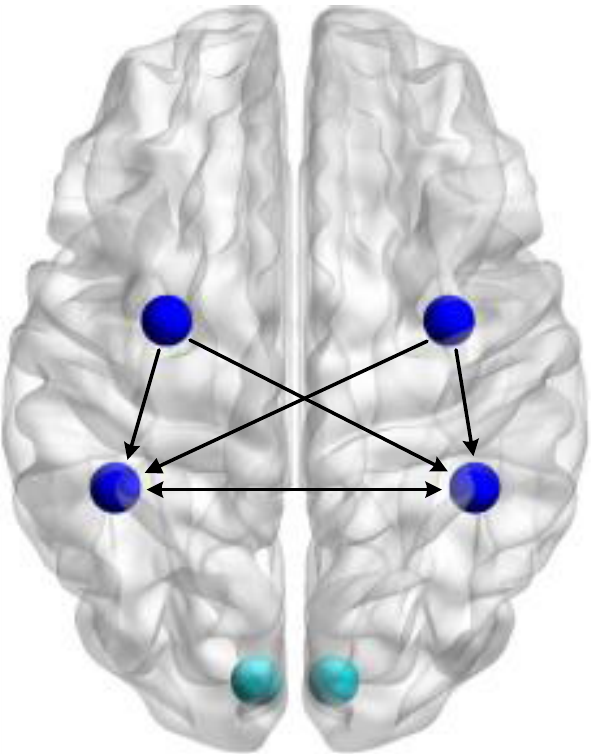}
		\end{minipage}%
	}
	\small 
	\centering
	\subfloat[MDL method (auditory)]{
		\begin{minipage}[t]{0.5\linewidth}	
			\centering
			\includegraphics[scale=0.45]{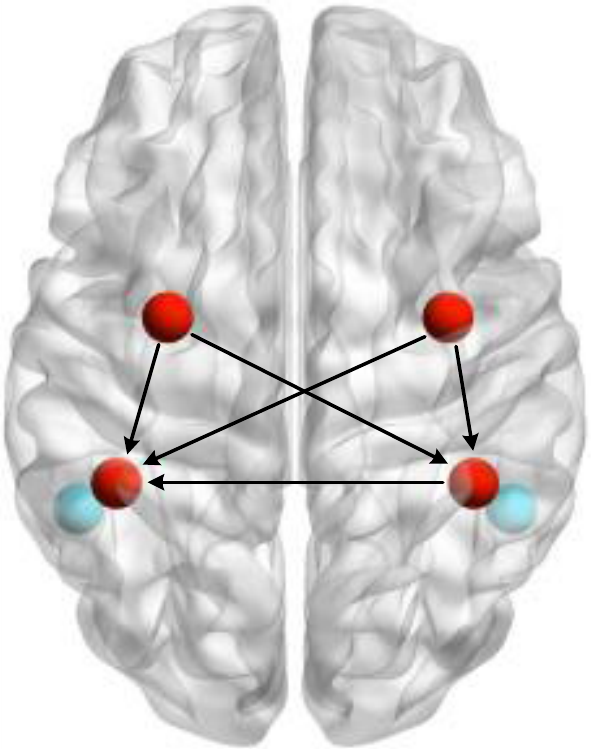}
		\end{minipage}%
	}
	\caption{Causal connectivities of subject 7 in mental calculation network under two stimuli. Cyan node represented visual/auditory stimulus node. The blue/red nodes related to the mental calculation network stated in the text. (a) and (b): mental calculation network obtained by conventional GCA. (c) and (d): mental calculation network obtained by proposed approach. }
	\label{fig:subject8_mental calculation network}	
\end{figure*}
Then, causal mental calculation networks of two subjects showed in Fig. \ref{fig:subject2_mental calculation network} and Fig. \ref{fig:subject8_mental calculation network} respectively. As stated above, the input node of stimulus should not be included into the network to be compared, we removed the causal connection between stimuli nodes and other four nodes of mental calculation network. Obviously, our proposal also had a desirable robustness in fMRI data. As seen in Fig. \ref{fig:subject2_mental calculation network} and Fig. \ref{fig:subject8_mental calculation network}, for subject 2 and subject 7, the mental calculation networks under different stimuli were almost identical. By the way, even for different subjects, the causal connection networks were similar at a large extent. As for conventional two-stage GCA, causal connection networks of two subjects above were more irregular, which seen in Fig. \ref{fig:subject2_mental calculation network} and Fig. \ref{fig:subject8_mental calculation network}. Comparing the similarity between causal networks identified by our proposal, the networks obtained by conventional GCA had almost no consistency characteristics, the networks under different stimuli appeared to be unrelated. Conventional two-stage GCA identified a inconclusive results in mental calculation. Consistent with the results in simulation, our proposal was more robust in identifying causal connection network, especially in complex networks. 

\section{Conclusions and Perspective}
\label{sec:conclusion}

\subsection{Contributions and Discussions}

The novelty of the present study is not in including the MDL principle in the GCA procedure, but rather in considering the MDL principle as unity strategy for model selection in the analysis procedure. Conventional GCA usually consists of two stages: ($1$) AIC/BIC for the predictors associated with internal or external information, and ($2$) $F$-statistic for evaluating the relative effects of exogenous variables. We emphasize that these two parts fall within the scope of model selection, and distinct theory might generate different model selection criteria. Model selection should follow the same mathematical theory in the GCA process. 

In this paper, we have addressed this concern and have proposed a unified model selection approach based on the MDL principle for GCA in the context of the general regression model paradigm. We have demonstrated its efficacy over conventional two-stage GCA approach in a $3$-node network and a $5$-node network synthetic experiments. All results confirm the superiority of proposed approach over conventional two-stage GCA. The unified model selection approach is capable of identifying the actual connectivity in all the cases, at the same time, avoiding greatly the false influences caused by the presence of noise. GCA was originally designed to handle pairs of variables, and may produce misleading results when the true relationship involves three or more variables \cite{Granger1969Investigating,Granger_JE74}. This case occurs when single connections are strong enough: if $A \rightarrow B$ and $B \rightarrow C$, then very likely $A \rightarrow C$ will be picked up by GCA, thus inducing a denser connected network than the truth. Our results suggest that the proposed approach can remove the indirect connection and retain the direct connection. The use of a better model selection strategy can improve the performance of GCA in systems involving a lager number of variables. As noted earlier (section \ref{sec:problem statement}), for the statistical strategy, the resulting network are heavily dependent on the significance level chosen. In the simulation experiments involving a $5$-node system, while a moderate significance level ($P < 0.05$) produces higher sensitivity in revealing actual connectivity (true positive, TP), it has lower specificity in avoiding false connectivity (true negative, TN). A more stringent significance threshold ($P < 0.01$) makes the reverse effect. In contrast, the proposed approach maintains high sensitivity and specificity in all cases, indicating that the MDL principle do much more than a higher significant level in F-statistics. The results illustrate the benefits of performing comparisons over all competing models in the model space as well.

More importantly, the proposed approach obtained more consistent results in a challenging fMRI dataset, in where visual/auditory stimuli with the same presentation design give identical neural correlates of mental calculation, allowing one to evaluate the performance of different GCA methods. This provides clear experimental proof that unified GCA is superior to conventional GCA \cite{li2020neural}. The results are also consistent with the current consensus that model selection is crucial to investigate causal connectivity using neuroimaging techniques \cite{valdes2011effective}. 

Essentially, both the MDL and statistical strategies, indeed all model selection methods, attempt to seek a trade-off between goodness-of-fit and complexity of the model involved. The MDL uses the code length which describes both the model complexity and the fitting error to achieve such a trade-off, whereas the statistical strategy implicitly conveys it in F-distribution function through degrees of freedom. In the MDL framework all competitive models can directly compare in terms of description length, without the need for an intermediate model. In contrast, F-statistics resorts the extra sum of squares principle to determine which model gives a better fit to the data. This approach requires one model (the restricted model) to be nested within another model (the unrestricted model). We argue that the pairwise comparison in F-statistics is responsible for the inferior performances when using the statistical strategy. The procedure of model selection is confined to comparisons between models with a nested relationship, thereby impairing the performance of the statistical strategy.

\subsection{Future Works}

In this study we have focused on a general regression model paradigm that provides a natural solution from two part form of MDL to the GCA. As a general principle for statistical model selection, the MDL principle develops many forms of description length in terms of coding schemes. The MDL forms can be viewed as imposing an adaptive penalty on model size. Although all forms achieve pointwise and minimax lower bounds on redundancy \cite{Hansen2001Model}, further investigation is required to determine the optimal coding scheme suitable for a given neuroimaging modality or noise level. Future animal experiments that provide intracranial recordings of and fMRI measurements of neural network synchronously  on the same animal will help to direct and test the development of these forms.	
Moreover, the GCA schemes were also conveyed in different function spaces \cite{Marinazzo_PRL08,angelini2009granger}. The MDL  will provide potential approaches to the Granger-causality representations in other function spaces principle because has rich connections with Bayesian statistics \cite{Hansen2001Model}.  More importantly, the MDL principle allows the comparison between any two model classes in terms of code length, regardless of their forms \cite{Hansen2001Model}. This robust feature has potential to accommodate the representations of Granger-causality generated from different model classes, while the causal correlation can be also investigated between disparate function spaces.

\bibliography{refs}

\begin{thebibliography}{10}
\providecommand{\url}[1]{#1}
\csname url@samestyle\endcsname
\providecommand{\newblock}{\relax}
\providecommand{\bibinfo}[2]{#2}
\providecommand{\BIBentrySTDinterwordspacing}{\spaceskip=0pt\relax}
\providecommand{\BIBentryALTinterwordstretchfactor}{4}
\providecommand{\BIBentryALTinterwordspacing}{\spaceskip=\fontdimen2\font plus
\BIBentryALTinterwordstretchfactor\fontdimen3\font minus
  \fontdimen4\font\relax}
\providecommand{\BIBforeignlanguage}[2]{{%
\expandafter\ifx\csname l@#1\endcsname\relax
\typeout{** WARNING: IEEEtran.bst: No hyphenation pattern has been}%
\typeout{** loaded for the language `#1'. Using the pattern for}%
\typeout{** the default language instead.}%
\else
\language=\csname l@#1\endcsname
\fi
#2}}
\providecommand{\BIBdecl}{\relax}
\BIBdecl

\bibitem{Amunts2016The}
K.~Amunts, C.~Ebell, J.~Muller, M.~Telefont, A.~Knoll, and T.~Lippert, ``The
  human brain project: Creating a european research infrastructure to decode
  the human brain,'' \emph{Neuron}, vol.~92, no.~3, pp. 574--581, 2016.

\bibitem{Martin2016The}
C.~L. Martin and M.~Chun, ``The brain initiative: Building, strengthening, and
  sustaining,'' \emph{Neuron}, vol.~92, no.~3, pp. 570--573, 2016.

\bibitem{Poo2016China}
M.~M. Poo, J.~L. Du, N.~Y. Ip, Z.~Q. Xiong, B.~Xu, and T.~Tan, ``China brain
  project: Basic neuroscience, brain diseases, and brain-inspired computing,''
  \emph{Neuron}, vol.~92, no.~3, pp. 591--596, 2016.

\bibitem{Committee2016Australian}
A.~B. A.~S. Committee \emph{et~al.}, ``Australian brain alliance,''
  \emph{Neuron}, vol.~92, no.~3, pp. 597--600, 2016.

\bibitem{Okano2016Brain}
H.~Okano, E.~Sasaki, T.~Yamamori, A.~Iriki, T.~Shimogori, Y.~Yamaguchi,
  K.~Kasai, and A.~Miyawaki, ``Brain/minds: A japanese national brain project
  for marmoset neuroscience,'' \emph{Neuron}, vol.~92, no.~3, pp. 582--590,
  2016.

\bibitem{Jeong2016Korea}
S.~J. Jeong, H.~Lee, E.~M. Hur, Y.~Choe, J.~W. Koo, J.~C. Rah, K.~J. Lee, H.~H.
  Lim, W.~Sun, and C.~Moon, ``Korea brain initiative: Integration and control
  of brain functions,'' \emph{Neuron}, vol.~92, no.~3, pp. 607--611, 2016.

\bibitem{strogatz2001exploring}
S.~H. Strogatz, ``Exploring complex networks,'' \emph{nature}, vol. 410, no.
  6825, p. 268, 2001.

\bibitem{Sporns2004Organization}
O.~Sporns, D.~R. Chialvo, M.~Kaiser, and C.~C. Hilgetag, ``Organization,
  development and function of complex brain networks,'' \emph{Trends in
  Cognitive Sciences}, vol.~8, no.~9, pp. 418--425, 2004.

\bibitem{Bullmore2009Complex}
E.~Bullmore and O.~Sporns, ``Complex brain networks: graph theoretical analysis
  of structural and functional systems,'' \emph{Nature Reviews Neuroscience},
  vol.~10, no.~3, pp. 186--198, 2009.

\bibitem{bressler2010large}
S.~L. Bressler and V.~Menon, ``Large-scale brain networks in cognition:
  emerging methods and principles,'' \emph{Trends in cognitive sciences},
  vol.~14, no.~6, pp. 277--290, 2010.

\bibitem{Rubinov2010Complex}
M.~Rubinov and O.~Sporns, ``Complex network measures of brain connectivity:
  Uses and interpretations,'' \emph{Neuroimage}, vol.~52, no.~3, pp.
  1059--1069, 2010.

\bibitem{sporns2011human}
O.~Sporns, ``The human connectome: a complex network,'' \emph{Annals of the New
  York Academy of Sciences}, vol. 1224, no.~1, pp. 109--125, 2011.

\bibitem{Siegel2012Spectral}
M.~Siegel, T.~H. Donner, and A.~K. Engel, ``Spectral fingerprints of
  large-scale neuronal interactions,'' \emph{Nature Reviews Neuroscience},
  vol.~13, no.~2, pp. 121--134, 2012.

\bibitem{Hutchison_NeuroImage13}
R.~M. Hutchison, T.~Womelsdorf, E.~A. Allen, P.~A. Bandettin, V.~D. Calhoun,
  M.~Corbetta, S.~D. Penna, J.~H. Duyn, G.~H. Glover, J.~Gonzalez-Castillo,
  D.~A. Handwerker, S.~Keiholz, V.~Kiviniemi, D.~A. Leopold, F.~d.~Pasquale,
  O.~Sporns, M.~Walter, and C.~Chang, ``{Dynamic Functional Connectivity:
  Promise, Issues, and Interpretations.}'' \emph{NeuroImage}, vol.~80, pp.
  360--378, 2013.

\bibitem{Valk2017Socio}
S.~L. Valk, B.~C. Bernhardt, A.~Böckler, M.~Trautwein, and T.~Singer,
  ``Socio-cognitive phenotypes differentially modulate large-scale structural
  covariance networks,'' \emph{Cerebral Cortex}, vol.~27, no.~2, 2017.

\bibitem{Stephan_NeuroImage12}
K.~E. Stephan and A.~Roebroeck, ``{A Short History of Causal Modeling of fMRI
  Data},'' \emph{NeuroImage}, vol.~62, pp. 856--863, 2012.

\bibitem{Tsunada_NN16}
J.~Tsunada, A.~S.~K. Liu, J.~I. Gold, and Y.~E. Cohen, ``Causal contribution of
  primate auditory cortex to auditory perceptual decision-making,''
  \emph{Nature Neuroscience}, vol.~19, pp. 135--142, 2016.

\bibitem{Olivier2008Identifying}
D.~Olivier, G.~Isabelle, S.~Sandrine, R.~Sebastien, D.~Colin, S.~Christoph,
  D.~Antoine, and V.-S. Pedro, ``Identifying neural drivers with functional
  mri: An electrophysiological validation,'' \emph{Plos Biology}, vol.~6,
  no.~12, pp. 2683--97, 2008.

\bibitem{van2014alpha}
T.~Van~Kerkoerle, M.~W. Self, B.~Dagnino, M.-A. Gariel-Mathis, J.~Poort, C.~Van
  Der~Togt, and P.~R. Roelfsema, ``Alpha and gamma oscillations characterize
  feedback and feedforward processing in monkey visual cortex.''
  \emph{Proceedings of the National Academy of Sciences}, vol. 111, no.~40, pp.
  14\,332--14\,341, 2014.

\bibitem{Bastos2015Visual}
A.~Bastos, J.~Vezoli, C.~A. Bosman, J.~M. Schoffelen, R.~Oostenveld, J.~R.
  Dowdall, P.~Deweerd, H.~Kennedy, and P.~Fries, ``Visual areas exert
  feedforward and feedback influences through distinct frequency channels.''
  \emph{Neuron}, vol.~85, no.~2, pp. 390--401, 2015.

\bibitem{brovelli2004beta}
A.~Brovelli, M.~Ding, A.~Ledberg, Y.~Chen, R.~Nakamura, and S.~L. Bressler,
  ``Beta oscillations in a large-scale sensorimotor cortical network:
  directional influences revealed by granger causality,'' \emph{Proceedings of
  the National Academy of Sciences}, vol. 101, no.~26, pp. 9849--9854, 2004.

\bibitem{Wang2008Estimating}
X.~Wang, Y.~Chen, and M.~Ding, ``Estimating granger causality after stimulus
  onset: A cautionary note,'' \emph{Neuroimage}, vol.~41, no.~3, pp. 767--776,
  2008.

\bibitem{gow2008lexical}
D.~W. Gow~Jr, J.~A. Segawa, S.~P. Ahlfors, and F.-H. Lin, ``Lexical influences
  on speech perception: a granger causality analysis of meg and eeg source
  estimates,'' \emph{Neuroimage}, vol.~43, no.~3, pp. 614--623, 2008.

\bibitem{Ploner2010Functional}
M.~Ploner, J.~M. Schoffelen, A.~Schnitzler, and J.~Gross, ``Functional
  integration within the human pain system as revealed by granger causality,''
  \emph{Human Brain Mapping}, vol.~30, no.~12, pp. 4025--4032, 2010.

\bibitem{florin2010effect}
E.~Florin, J.~Gross, J.~Pfeifer, G.~R. Fink, and L.~Timmermann, ``The effect of
  filtering on granger causality based multivariate causality measures,''
  \emph{Neuroimage}, vol.~50, no.~2, pp. 577--588, 2010.

\bibitem{Roebroeck2005Mapping}
A.~Roebroeck, E.~Formisano, and R.~Goebel, ``Mapping directed influence over
  the brain using granger causality and fmri,'' \emph{Neuroimage}, vol.~25,
  no.~1, pp. 230--242, 2005.

\bibitem{sridharan2008critical}
D.~Sridharan, D.~J. Levitin, and V.~Menon, ``A critical role for the right
  fronto-insular cortex in switching between central-executive and default-mode
  networks,'' \emph{Proceedings of the National Academy of Sciences}, vol. 105,
  no.~34, pp. 12\,569--12\,574, 2008.

\bibitem{Menon2010saliency}
V.~Menon and L.~Q. Uddin, ``Saliency, switching, attention and control: A
  network model of insula function,'' \emph{Brain Structure and Function}, vol.
  214, no. 5-6, pp. 655--667, 2010.

\bibitem{Ryali2011Multivariate}
S.~Ryali, K.~Supekar, T.~Chen, and V.~Menon, ``Multivariate dynamical systems
  models for estimating causal interactions in fmri,'' \emph{Neuroimage},
  vol.~54, no.~2, pp. 807--823, 2011.

\bibitem{wiener1956theory}
N.~Wiener, ``The theory of prediction.'' \emph{Modern mathematics for
  engineers}, 1956.

\bibitem{Granger1969Investigating}
C.~W.~J. Granger, ``Investigating causal relations by econometric models and
  cross-spectral methods.'' \emph{Econometrica}, vol.~37, no.~3, pp. 424--438,
  1969.

\bibitem{Granger_JE74}
C.~Granger and P.~Newbold, ``{Spurious Regressions in Econometrics.}''
  \emph{Journal of Econometrics}, vol.~2, pp. 111--120, 1974.

\bibitem{chavez2003statistical}
M.~Ch{\'a}vez, J.~Martinerie, and M.~Le~Van~Quyen, ``Statistical assessment of
  nonlinear causality: application to epileptic eeg signals,'' \emph{Journal of
  neuroscience methods}, vol. 124, no.~2, pp. 113--128, 2003.

\bibitem{gourevitch2006linear}
B.~Gour{\'e}vitch, R.~Le~Bouquin-Jeann{\`e}s, and G.~Faucon, ``Linear and
  nonlinear causality between signals: methods, examples and neurophysiological
  applications,'' \emph{Biological cybernetics}, vol.~95, no.~4, pp. 349--369,
  2006.

\bibitem{Goebel_MRI03}
R.~Goebel, A.~Roebroeck, D.~S. Kim, and E.~Formisano, ``{Investigating Directed
  Cortical Interactions in Time-Resolved fMRI Data Using Vector Autoregressive
  Nodeling and Granger Causality Mapping.}'' \emph{Magnetic Resonance Imaing},
  vol.~21, pp. 1251--1261, 2003.

\bibitem{Hacker_AE06}
R.~S. Hacker and A.~Hatemi-J, ``{Tests for Causality Between Integrated
  Variables Using Asymptotic and Bootstrap Distributions: Theory and
  Application.}'' \emph{Applied Economics}, vol.~38, no.~13, pp. 1489--1500,
  2006.

\bibitem{Kaminski}
M.~Kaminski, \emph{Multichannel Data Analysis in Biomedical Research.},
  2nd~ed.\hskip 1em plus 0.5em minus 0.4em\relax Pacific Grove, CA, USA:
  Duxbury Resource Center, 2007.

\bibitem{Hatemi-J_EE12}
A.~Hatemi-J, ``{Asymmetric Causality Tests with an Application.}''
  \emph{Empirical Economics}, vol.~43, no.~1, pp. 447--456, 2012.

\bibitem{John1982Measurement}
J.~Geweke, ``Measurement of linear dependence and feedback between multiple
  time series,'' \emph{Publications of the American Statistical Association},
  vol.~77, no. 378, pp. 304--313, 1982.

\bibitem{Ding2006Granger}
M.~Ding, Y.~Chen, and S.~L. Bressler, ``Granger causality: Basic theory and
  application to neuroscience,'' \emph{Quantitative Biology}, pp. 826--831,
  2006.

\bibitem{Guo2008Uncovering}
S.~Guo, J.~Wu, M.~Ding, and J.~Feng, ``Uncovering interactions in the frequency
  domain,'' \emph{Plos Computational Biology}, vol.~4, no.~5, p. e1000087,
  2008.

\bibitem{seth2010matlab}
A.~K. Seth, ``A matlab toolbox for granger causal connectivity analysis,''
  \emph{Journal of neuroscience methods}, vol. 186, no.~2, pp. 262--273, 2010.

\bibitem{Barnett2014The}
L.~Barnett and A.~K. Seth, ``The mvgc multivariate granger causality toolbox: A
  new approach to granger-causal inference,'' \emph{Journal of Neuroscience
  Methods}, vol. 223, pp. 50--68, 2014.

\bibitem{Stokes2017A}
P.~A. Stokes and P.~L. Purdon, ``A study of problems encountered in granger
  causality analysis from a neuroscience perspective,'' \emph{Proc Natl Acad
  Sci U S A}, vol. 114, no.~34, p. E7063, 2017.

\bibitem{ning2017dynamic}
L.~Ning and Y.~Rathi, ``A dynamic regression approach for frequency-domain
  partial coherence and causality analysis of functional brain networks,''
  \emph{IEEE transactions on medical imaging}, vol.~37, no.~9, pp. 1957--1969,
  2017.

\bibitem{liao2009kernel}
W.~Liao, D.~Marinazzo, Z.~Pan, Q.~Gong, and H.~Chen, ``Kernel granger causality
  mapping effective connectivity on fmri data,'' \emph{IEEE transactions on
  medical imaging}, vol.~28, no.~11, pp. 1825--1835, 2009.

\bibitem{Chen_NC14}
Z.~T. Chen, K.~Zhang, L.~W. Chan, and B.~Sch\"{o}lkopf, ``Causal discovery via
  reproducing kernel hibert space embedding,'' \emph{Neural Computation},
  vol.~26, no.~7, pp. 1484--1517, 2014.

\bibitem{Newman_PRE04}
M.~E.~J. Newman and M.~Girvan, ``{Finding and evaluating community structure in
  networks.}'' \emph{Physical Review E}, vol.~69, p. {026113}, 2004.

\bibitem{Marinazzo_PRL08}
D.~Marinazzo, M.~Pellicoro, and S.~Stramaglia, ``{Kernel Method for Nonlinear
  Granger Causality.}'' \emph{Physical Review Letters}, vol. 100, no.~14, p.
  144103, 2008.

\bibitem{Zou_BMCB10}
C.~L. Zou, C.~Ladroue, S.~X. Guo, and J.~F. Feng, ``Identifying interactions in
  the time and frequency domains in local and global networks - a granger
  causality approach.'' \emph{BMC Bioinformatics}, vol. 337, no.~11, pp. 1--17,
  2010.

\bibitem{Akaike1974new}
H.~Akaike, ``A new look at the statistical model identification.'' in
  \emph{Selected Papers of Hirotugu Akaike}.\hskip 1em plus 0.5em minus
  0.4em\relax Springer, 1974, pp. 215--222.

\bibitem{Schwarz1978estimating}
G.~Schwarz \emph{et~al.}, ``Estimating the dimension of a model.'' \emph{The
  annals of statistics}, vol.~6, no.~2, pp. 461--464, 1978.

\bibitem{Deshpande2010Multivariate}
G.~Deshpande, S.~Laconte, G.~A. James, S.~Peltier, and X.~Hu, ``Multivariate
  granger causality analysis of fmri data.'' \emph{Human Brain Mapping},
  vol.~30, no.~4, pp. 1361--1373, 2010.

\bibitem{Guo2010Granger}
S.~Guo, C.~Ladroue, and J.~Feng, \emph{Granger Causality: Theory and
  Applications}.\hskip 1em plus 0.5em minus 0.4em\relax Springer, 2010.

\bibitem{Bressler2011Wiener}
S.~L. Bressler and A.~K. Seth, ``Wiener–granger causality: A well established
  methodology,'' \emph{Neuroimage}, vol.~58, no.~2, pp. 323--329, 2011.

\bibitem{Li2010A}
X.~Li, G.~Marrelec, R.~F. Hess, and H.~Benali, ``A nonlinear identification
  method to study effective connectivity in functional mri,'' \emph{Medical
  Image Analysis}, vol.~14, no.~1, pp. 30--38, 2010.

\bibitem{Casella}
G.~Casella and R.~L. Berger, \emph{{Statistical Inference}}, 2nd~ed.\hskip 1em
  plus 0.5em minus 0.4em\relax Pacific Grove, CA, USA: Duxbury Resource Center,
  2001.

\bibitem{wasserstein2016asa}
R.~L. Wasserstein, N.~A. Lazar \emph{et~al.}, ``The asa’s statement on
  p-values: context, process, and purpose,'' \emph{The American Statistician},
  vol.~70, no.~2, pp. 129--133, 2016.

\bibitem{wasserstein2019moving}
R.~L. Wasserstein, A.~L. Schirm, and N.~A. Lazar, ``Moving to a world beyond
  “p< 0.05”,'' 2019.

\bibitem{amrhein2019scientists}
V.~Amrhein, S.~Greenland, and B.~McShane, ``Scientists rise up against
  statistical significance,'' 2019.

\bibitem{mcshane2019abandon}
B.~B. McShane, D.~Gal, A.~Gelman, C.~Robert, and J.~L. Tackett, ``Abandon
  statistical significance,'' \emph{The American Statistician}, vol.~73, no.
  sup1, pp. 235--245, 2019.

\bibitem{amrhein2018remove}
V.~Amrhein and S.~Greenland, ``Remove, rather than redefine, statistical
  significance,'' \emph{Nature Human Behaviour}, vol.~2, no.~1, p.~4, 2018.

\bibitem{amrhein2019inferential}
V.~Amrhein, D.~Trafimow, and S.~Greenland, ``Inferential statistics as
  descriptive statistics: There is no replication crisis if we don’t expect
  replication,'' \emph{The American Statistician}, vol.~73, no. sup1, pp.
  262--270, 2019.

\bibitem{hurlbert2019coup}
S.~H. Hurlbert, R.~A. Levine, and J.~Utts, ``Coup de gr{\^a}ce for a tough old
  bull:“statistically significant” expires,'' \emph{The American
  Statistician}, vol.~73, no. sup1, pp. 352--357, 2019.

\bibitem{miao2011altered}
X.~Miao, X.~Wu, R.~Li, K.~Chen, and L.~Yao, ``Altered connectivity pattern of
  hubs in default-mode network with alzheimer's disease: an granger causality
  modeling approach,'' \emph{PloS one}, vol.~6, no.~10, p. e25546, 2011.

\bibitem{deshpande2011instantaneous}
G.~Deshpande, P.~Santhanam, and X.~Hu, ``Instantaneous and causal connectivity
  in resting state brain networks derived from functional mri data,''
  \emph{Neuroimage}, vol.~54, no.~2, pp. 1043--1052, 2011.

\bibitem{Li_TMI11}
X.~F. Li, D.~Coyle, L.~Maguire, T.~M. McGinnity, and H.~Benali, ``A model
  selection method for nonlinear system identification based fmri effective
  connectivity analysis.'' \emph{IEEE Transactions on Medical Imaging},
  vol.~30, no.~7, pp. 1365--1380, 2011.

\bibitem{Rissanen1978Modeling}
J.~Rissanen, ``Modeling by shortest data description,'' \emph{Automatica},
  vol.~14, no.~5, pp. 465--471, 1978.

\bibitem{bryant2000model}
P.~G. Bryant and O.~I. Cordero-Brana, ``Model selection using the minimum
  description length principle,'' \emph{The American Statistician}, vol.~54,
  no.~4, pp. 257--268, 2000.

\bibitem{Grunwald}
P.~D. Gr\"{u}nwald, I.~J. Myung, and M.~A. Pitt, \emph{Advances in Minimum
  Description Length: Theory and Applications}, ser. Neural Information
  Processing series.\hskip 1em plus 0.5em minus 0.4em\relax Cambridge,
  Massachusetts: The MIT Press, 2005.

\bibitem{Hansen_DECI99}
M.~Hansen and B.~Yu, ``Bridging aic and bic: An mdl model selection
  criterion.'' in \emph{Worskhop on Detection, Estimation, Classification, and
  Imaging}, Santa Fe, NM, USA, 1999, pp. 24--26.

\bibitem{Hansen2001Model}
M.~H. Hansen and B.~Yu, ``Model selection and the principle of minimum
  description length,'' \emph{Publications of the American Statistical
  Association}, vol.~96, no. 454, pp. 746--774, 2001.

\bibitem{Li2008introduction}
M.~Li, P.~Vit{\'a}nyi \emph{et~al.}, \emph{An introduction to Kolmogorov
  complexity and its applications.}\hskip 1em plus 0.5em minus 0.4em\relax
  Springer, 2008, vol.~3.

\bibitem{shannon1948mathematical}
C.~E. Shannon, ``A mathematical theory of communication,'' \emph{Bell system
  technical journal}, vol.~27, no.~3, pp. 379--423, 1948.

\bibitem{Rissanen1983A}
J.~Rissanen, ``A universal prior for integers and estimation by minimum
  description length,'' \emph{Annals of Statistics}, vol.~11, no.~2, pp.
  416--431, 1983.

\bibitem{rissanen1984universal}
------, ``Universal coding, information, prediction, and estimation,''
  \emph{IEEE Transactions on Information theory}, vol.~30, no.~4, pp. 629--636,
  1984.

\bibitem{Rissanen1986Stochastic}
------, ``Stochastic complexity and modeling,'' \emph{Annals of Statistics},
  vol.~14, no.~3, pp. 1080--1100, 1986.

\bibitem{Rissanen1987Stochastic}
------, ``Stochastic complexity,'' \emph{Journal of the Royal Statistical
  Society}, vol.~49, no.~3, pp. 223--239, 1987.

\bibitem{rissanen1996fisher}
J.~J. Rissanen, ``Fisher information and stochastic complexity,'' \emph{IEEE
  transactions on information theory}, vol.~42, no.~1, pp. 40--47, 1996.

\bibitem{kolmogorov1965three}
A.~N. Kolmogorov, ``Three approaches to the quantitative definition of
  information,'' \emph{Problems of information transmission}, vol.~1, no.~1,
  pp. 1--7, 1965.

\bibitem{kolmogorov1968logical}
A.~Kolmogorov, ``Logical basis for information theory and probability theory.''
  \emph{IEEE Transactions on Information Theory}, vol.~14, no.~5, pp. 662--664,
  1968.

\bibitem{Wallace1968An}
C.~S. Wallace and D.~M. Boulton, ``An information measure for classification,''
  \emph{The Computer Journal}, vol.~11, no.~2, pp. 185--194, 1968.

\bibitem{Rissanen}
J.~Rissanen, \emph{Information and Complexity in Statistical Modeling}, ser.
  Information Science and Statistics.\hskip 1em plus 0.5em minus 0.4em\relax
  New York, NY, USA: Springer Verlag, 2005.

\bibitem{Cover2012elements}
T.~M. Cover and J.~A. Thomas, \emph{Elements of information theory.}\hskip 1em
  plus 0.5em minus 0.4em\relax New York:Wiley, 1991.

\bibitem{grunwald2007minimum}
P.~D. Gr{\"u}nwald, \emph{The minimum description length principle}.\hskip 1em
  plus 0.5em minus 0.4em\relax MIT press, 2007.

\bibitem{Rissanen1989stochastic}
J.~Rissanen, \emph{Stochastic complexity in statistical inquiry}.\hskip 1em
  plus 0.5em minus 0.4em\relax World Scientific, 1989.

\bibitem{li2020neural}
F.~Li, X.~Wang, P.~Shi, Q.~Lin, and Z.~Hu, ``Neural network can be revealed
  with functional mri: evidence from self-consistent experiment,'' \emph{Nature
  Communications (Submitted)}, 2020.

\bibitem{valdes2011effective}
P.~A. Valdes-Sosa, A.~Roebroeck, J.~Daunizeau, and K.~Friston, ``Effective
  connectivity: influence, causality and biophysical modeling,''
  \emph{Neuroimage}, vol.~58, no.~2, pp. 339--361, 2011.

\bibitem{angelini2009granger}
L.~Angelini, M.~Pellicoro, and S.~Stramaglia, ``Granger causality for circular
  variables,'' \emph{Physics Letters A}, vol. 373, no.~29, pp. 2467--2470,
  2009.

\end{thebibliography}
\bibliographystyle{IEEEtran}

\EOD

\end{document}